\documentclass[prb,unsortedaddress,superscriptaddress,twocolumn,showpacs,floatfix,nofootinbib]{revtex4-1}
\usepackage{graphicx}
\usepackage{times}
\usepackage{amsmath,amssymb}
\usepackage{multirow} 

\allowdisplaybreaks[4]
\newcommand{\ket}[1]{\left| #1 \right\rangle}
\newcommand{\id}{{\bf 1}}
\newcommand{\btau}{{\boldsymbol{\tau}}}
\newcommand{\ch}{{\rm ch}}

\begin{document}

\title{Microscopic models of interacting Yang-Lee anyons}

\author{E. Ardonne}
\affiliation{Nordita, Roslagstullsbacken 23, 106-91 Stockholm, Sweden}

\author{J. Gukelberger}
\affiliation{Theoretische Physik, ETH Zurich, 8093 Zurich, Switzerland}

\author{A.W.W. Ludwig}
\affiliation{Physics Department, University of California, Santa Barbara, CA 93106, USA}

\author{S. Trebst}
\affiliation{Microsoft Research, Station Q,
University of California, Santa Barbara, CA 93106, USA} 

\author{M. Troyer}
\affiliation{Theoretische Physik, ETH Zurich, 8093 Zurich, Switzerland}

\date{\today}

\begin{abstract}
Collective states of interacting  non-Abelian anyons have recently
been studied mostly in the context of certain fractional quantum Hall states,
such as the Moore-Read state proposed to describe the physics of the quantum
Hall plateau at filling fraction $\nu = 5/2$.
In this manuscript, we further expand this line of research and present {\sl non-unitary} generalizations 
of interacting anyon models. In particular, we introduce the notion of
Yang-Lee anyons, discuss their relation to the so-called `Gaffnian' quantum
Hall wave function, and describe an elementary model for their
interactions. 
A one-dimensional version of this model 
-- a non-unitary generalization of the original golden chain model --
can be fully understood in terms of an exact algebraic solution and
numerical diagonalization. We discuss the gapless theories of these
chain models for general su(2)$_k$ anyonic theories and their Galois conjugates.
We further introduce and solve a one-dimensional version of the Levin-Wen model for 
non-unitary Yang-Lee anyons.
\end{abstract}


\maketitle


\section{Introduction}

Over thirty years ago Leinaas and Myrheim pointed out \cite{lm77}, that in systems confined to two spatial 
dimensions particles with exotic exchange statistics, more general than those of bosons and fermions,
are possible. Such particles with arbitrary exchange statistics were later coined {\em anyons} by Wilzeck~\cite{w82}. 
Today it is widely believed that this possibility is indeed realized in the fractional quantum Hall effect. 
An even more intriguing form of statistics has recently received considerable attention, namely that of non-Abelian statistics,
first proposed in a seminal paper by Moore and Read\cite{mr91}.
This form of statistics can occur in two-dimensional systems in which introducing excitations gives rise to a {\em macroscopic degeneracy}
of states. 
Upon braiding the excitations, the wave function 
(or better, the vector of wave functions) 
describing the system, does not merely acquire an overall phase, but can actually transform into one
another, as described by a unitary braid matrix acting within the degenerate manifold. In general, these braid matrices do not commute, hence
the name non-Abelian statistics. 
While several systems have been theoretically proposed to exhibit quasiparticles with non-Abelian statistics, 
such as unconventional $p_x + i p_y$ superconductors \cite{ReadGreen}, rotating Bose-Einstein condensates \cite{Cooper01},
 or certain heterostructures involving a novel class of materials, so-called topological insulators \cite{FuKanePplusIP}, 
 the system currently being cast under intense experimental scrutiny
is the fractional quantum Hall effect observed at filling fraction $\nu=5/2$, with some evidence suggesting that this
state is indeed non-Abelian in nature~\cite{wpw09,boi10}.

An early attempt to describe this $\nu=5/2$ quantum Hall state came in the form of the so-called
Haldane-Rezayi wave function\cite{hr88}, which can be thought of as a $d$-wave paired
BCS condensate of composite fermions -- electrons with, in this case, two flux quanta attached.
A peculiar feature of the Haldane-Rezayi state is that its gapless edge modes are described by a 
{\sl non-unitary} conformal field theory \cite{ww84,mr96,g93,gfn97,gl98} with central charge $c=-2$.
However, non-unitary dynamics cannot describe a physical system as it would
violate basic principles of quantum mechanics.
It has therefore been argued that in general the non-unitary nature of an edge state indicates that the 
underlying phase is not bulk gapped, but in fact critical \cite{r09b,r09a,Gaffnian} in which case the
edge states lose their identity as they dissipate into the gapless bulk.
For the Haldane-Rezayi state it indeed turns out that it does not describe a gapped topological
phase, but rather the gapless state at a quantum phase transition between two gapped states -- 
the Moore-Read quantum Hall state~\cite{mr91} and a so-called strong pairing quantum Hall phase~\cite{rg00}.
Since the work of Haldane and Rezayi many other wave functions have been proposed, 
which appear to have non-unitary edge state theories \cite{src07,bh08,es09,s3}.
One particularly well-known example is the so-called `Gaffnian state'\cite{Gaffnian}.
These proposed states typically arise from numerical studies of model Hamiltonians, for which
it is oftentimes hard to determine whether they describe a gapped or gapless phase.
The ultimate fate of these newly proposed states has therefore remained an active field
of research.

In this paper, we will study the excitations of the putative quantum liquid described by such non-unitary 
wave functions from an `anyon-model' perspective. 
Thus, we explore the physics that would occur if one were to assume that such excitations indeed exist
and study in detail their collective behavior that would result. We refer to these excitations as non-unitary anyons
and model the interaction between these anyons in the same way as it was done
in the unitary case \cite{Feiguin07}.
It will turn out that the Hamiltonians one is led to consider for the non-unitary anyons
are, not surprisingly, non-Hermitian. Nevertheless, the energy spectrum turns out to be completely real.
One can hope that the results of such a study would shed some light on the questions
surrounding the physical meaning of non-unitary anyons.
Of course, in the context of two-dimensional {\sl classical} statistical mechanics models
having a non-unitary conformal field theory (CFT) describing a critical phase is a situation which often  occurs
-- see, for instance, the (restricted) height models described in Ref. \onlinecite{fb84}, which are generalizations
of the `restricted solid-on-solid' models\cite{abf84} -- and is not problematic. 

Before closing this introduction with the outline of the paper, we should mention one additional
point, which motivates the study on non-unitary anyon models: the concept of
`Galois-conjugation' of conformal field theories. In short, the action of the Galois group
relates different CFT's which have the same number of fields, obeying the same fusion
rules. The non-unitary models we encounter in this paper, are in fact Galois-conjugates
of unitary CFT's, and thus, the study on non-unitary anyons is closely related to the study of
Galois conjugation in conformal field theory~\cite{dbg91,cg94}.
For the purposes of this paper we will however not need to use any details
of these connections.

This paper is organized as follows: 
In Sec.~\ref{Sec:YL-Chains} we discuss the Yang-Lee anyonic chains.  
We start by giving 
a review of the
 derivation of the Hamiltonian of the 
(ordinary, i.e. unitary)
golden chain \cite{Feiguin07} of Fibonacci 
anyons in Sec. \ref{Sec:GoldenChain}, and a  short mathematical description 
of the Galois-conjugated model, the 
so-called
Yang-Lee chain in Sec.~\ref{Sec:YL-Models}, including an explicit derivation of their non-Hermitian, microscopic Hamiltonians, and discuss their relation to the `golden chain' models of Ref.~\onlinecite{Feiguin07} 
via Galois conjugation. 
We continue in Sec. \ref{Sec:YL-AnalyticalSolution} with a discussion of the algebraic structure underlying these microscopic Hamiltonians, which in turn allows for an analytical identification of their gapless theories in terms of certain non-unitary minimal models of conformal field theory. We next present a number of exact, numerical results in Sec.~\ref{Sec:YL-Numerics}, and discuss the topological symmetry protecting the critical states in Sec. \ref{Sec:TopSym} before rounding off with a discussion of the case of general level $k$ in su(2)$_k$ theories in Sec. \ref{Sec:su2k}

In Sec.~\ref{Sec:YL-Ladders} we turn to `doubled Yang-Lee' models and, in particular, a non-unitary generalization of the 
(unitary) high-genus ladder model which was 
previously studied for Fibonacci anyons in Ref.~\onlinecite{Gils09b}, where we introduce the model in Sec. \ref{Sec:LadderHam}. We then discuss the phase diagram in Sec. \ref{Sec:PhaseDiag}, present the analytical solution at special critical points in Sec. \ref{Sec:CriticalPoints} and finish with exact numerical results in Sec. \ref{Sec:YL-ladder-num}.

The final section summarizes the results and discusses their relevance to hypothetical non-unitary topological phases. An appendix contains a detailed discussion of the conformal energy spectra at the four critical points discussed in the two models.


\section{Yang-Lee chains}
\label{Sec:YL-Chains}

At the focus of this manuscript are anyonic models that are certain {\sl non-unitary} generalizations 
of unitary non-Abelian anyon models, which have been extensively studied in the recent past 
\cite{Feiguin07,Bonesteel07,Trebst08a,Trebst08b,Fidkowski08,Fidkowski09,Gils09,Gils09b,Ludwig10}. 
The basic constituents of the generalizations considered here are non-unitary, non-Abelian anyons.
Like their unitary counterparts they carry a quantum number that corresponds to a generalized angular 
momentum in so-called su(2)$_k$ anyonic theories, which are certain deformations \cite{SU2q} of SU(2). 
We first concentrate on an elementary example where there is only a single anyon type
by explicitly considering the anyon theory su(2)$_3$. In the unitary version of this theory the elementary
degrees of freedom are often referred to as `Fibonacci anyons', and it is their non-unitary counterparts 
which we term `Yang-Lee anyons'.
We will return to a discussion of the non-unitary generalizations for general su(2)$_k$ anyonic theories
in Section~\ref{Sec:su2k}.

We start by quickly reviewing the basic construction of microscopic (chain) models of 
interacting non-Abelian anyons, following the ideas of the `golden chain' model of 
Ref.~\onlinecite{Feiguin07} and the detailed exposition of Ref.~\onlinecite{Trebst08b}.
The construction of these models proceeds in two steps. First, we describe the general
structure of the Hilbert space of these models in a particular `fusion chain' representation,
which is identical for the unitary and non-unitary models. In a second step we turn to
the microscopic form of the Hamiltonian capturing interactions between the anyons. 
While this second step is quite similar for the unitary and non-unitary cases,
the microscopic Hamiltonians for the two cases are distinct. 

\subsection{The golden chain}
\label{Sec:GoldenChain}

The elementary degrees of freedom in our microscopic model are the particle types
(or generalized angular momenta) of the su(2)$_3$ anyonic theory. 
In its simplest form (considering only integer momenta) this theory contains a trivial particle 
(or vacuum state), which we denote by $\bf 1$, and an anyonic particle, which we label as $\btau$. 
These particles can form combined states according to the fusion rules
\begin{align}
{\bf 1} \times {\bf 1} &= {\bf 1} & {\bf 1} \times \btau &= \btau & \btau \times \btau = {\bf 1} + \btau \ .
\label{Eq:FusionRules}
\end{align}
The non-Abelian nature of the anyonic $\btau$-particle reveals itself in the {\sl multiple} fusion 
outcomes when combining two of these particles.
\begin{figure}[b]
  \begin{center}
    \includegraphics[width=.5\columnwidth]{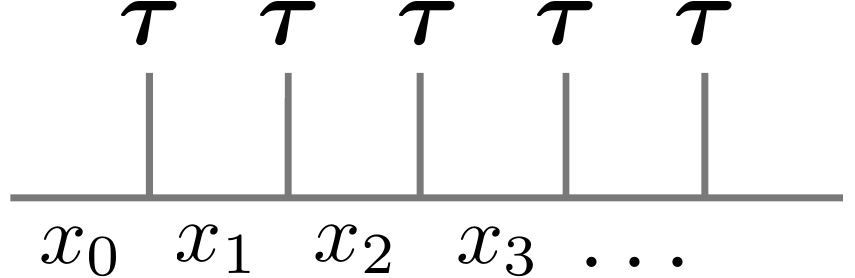}
  \end{center}
  \caption{A chain of Fibonacci or Yang-Lee anyons (denoted by the $\btau$'s in the upper row).
                The set of admissible labelings $\{x_i\}$ along the fusion chain (lines)
                constitutes the Hilbert space of the Yang-Lee (and Fibonacci) chains.}
\label{Fig:fusiontree}
\end{figure}
Our chain model then consists of $L$ such $\btau$-particles in a one-dimensional
arrangement as depicted on the top of Fig.~\ref{Fig:fusiontree},
where $L$ denotes the number of sites of the chain. 
Since pairs of $\btau$-particles can be fused into more than one state, such a system
of $L$ non-Abelian anyons spans a {\em macroscopic} manifold of states, 
i.e. a vector space whose dimension grows exponentially in the number of anyons. 
It is this manifold of states that constitutes the 
Hilbert space of our microscopic model.
To enumerate the states in the latter we define a so-called `fusion chain' as illustrated 
on the bottom of Fig.~\ref{Fig:fusiontree}. 
Here the original $\btau$-particles constituting 
the chain are denoted by the lines which are `incoming' from above. 
The links in the fusion chain carry labels $\{ x_i \}$ which again correspond to the particle 
types of the su(2)$_3$ theory. Reading the labels from left to right a labeling is called 
admissible if at each vertex the fusion rules \eqref{Eq:FusionRules} of su(2)$_3$ are obeyed, 
i.e. a  $\btau$ label is followed by either a $\bf 1$ or $\btau$ label, while a $\bf 1$ label is 
always followed by a $\btau$ label. Every such admissible labeling then constitutes one state
in the Hilbert space of our anyonic chain.
Considering periodic boundary conditions, i.e. $x_{L} = x_{0}$, it is straight forward to show that
the dimension of the Hilbert space is given in terms of Fibonacci numbers as 
\[
   \dim_{L} = {\rm Fib}_{L-1} + {\rm Fib}_{L+1} \,,
\]
where ${\rm Fib}_{i}$ denotes the $i$-th Fibonacci number, defined by 
${\rm Fib}_{i+1} = {\rm Fib}_{i} + {\rm Fib}_{i-1}$ and the initial conditions ${\rm Fib}_{1} = {\rm Fib}_{2} = 1$.

We now proceed to the second step of our construction, the derivation of a microscopic
Hamiltonian. In doing so we follow the perspective of the original `golden chain' model \cite{Feiguin07}
in assuming that interactions 
between a pair of neighboring $\btau$ particles 
-- mediated, for instance, by topological 
charge tunneling \cite{Bonderson09} -- will result in an energy splitting of the two possible 
fusion outcomes in Eq.~\eqref{Eq:FusionRules}. 
Our Hamiltonian captures 
this splitting
 by projecting the 
fusion outcome of two neighboring $\btau$ particles onto the 
trivial fusion channel, i.e. assigning an energy of $E_{\bf 1} = -1$ to the fusion of
two $\btau$ particles into the trivial channel and an energy of $E_{\btau} = 0$ to the
fusion into the $\btau$ channel.
This anyonic Hamiltonian is thus reminiscent of the common Heisenberg Hamiltonian for
SU(2) spins, which, for instance, projects two ordinary spin-1/2's onto the singlet channel and 
assigns a higher energy to the alternative triplet channel.

\begin{figure}[b]
\begin{center}
  \includegraphics[width=\columnwidth]{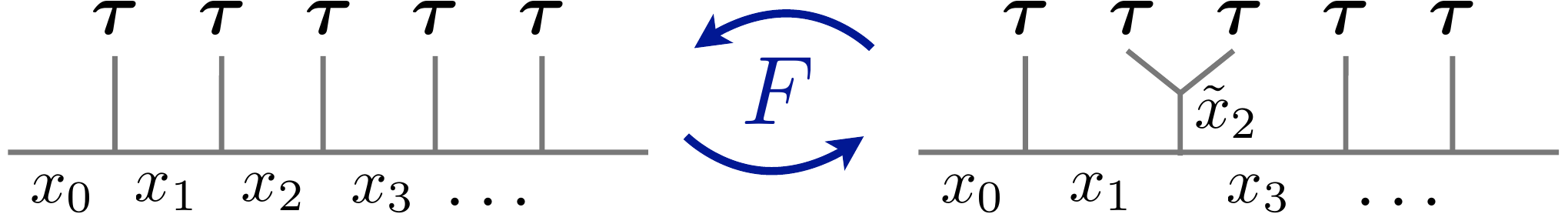}
\end{center}
\caption{The $F$-symbol describing the local change of basis.}
\label{Fig:Fsymbol}
\end{figure}

\begin{figure}[!ht]
\begin{center}
  \includegraphics[width=.8\columnwidth]{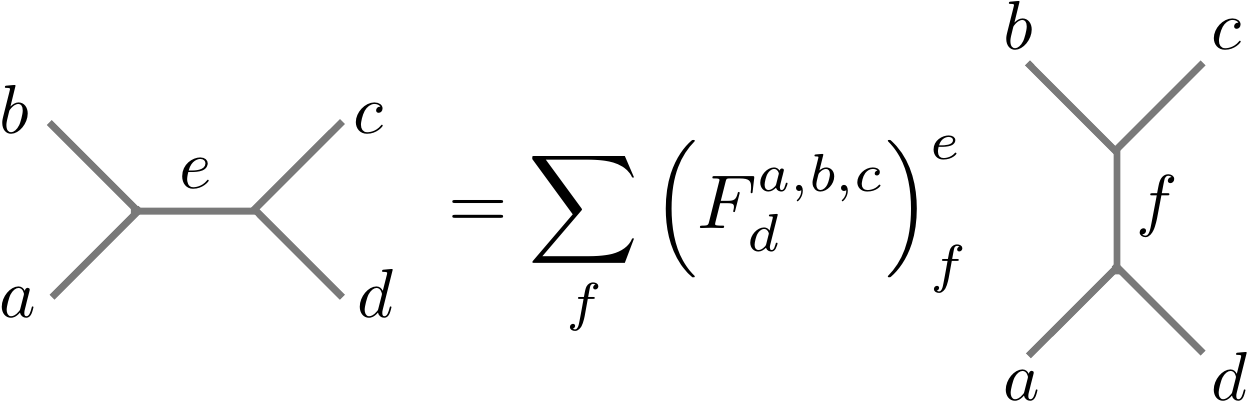}
\end{center}
\caption{The general form of the $F$-symbol.}
\label{Fig:Fgeneral}
\end{figure}

To explicitly derive the Hamiltonian in the Hilbert space of fusion chain labelings
introduced above, we note that in this basis the fusion of two neighboring $\btau$ 
particles is not explicit. To get direct access to this fusion channel of two neighboring
$\btau$ particles, we need to locally transform the basis as depicted in Fig.~\ref{Fig:Fsymbol}.
The matrix describing this transformation is typically called the $F$-symbol, which can
be thought of as an anyonic generalization of Wigner's $6j$-symbol for ordinary $SU(2)$ 
spins. 
Its general form (in the absence of fusion multiplicities) is given in 
Fig.~\ref{Fig:Fgeneral}.

Assuming that we know the explicit form of the $F$-symbols (see the next section for more details), 
we can now explicitly derive the microscopic Hamiltonian in the fusion chain basis. 
After the basis transformation, the fusion channel of the two neighboring
anyons is manifest, so by means of a simple projection we can assign an energy 
to each of the fusion channels. The final step left after this projection, is to transform back
to the original basis, which again employs the $F$-symbol.

To make the individual steps of this derivation more explicit, we consider the example of Fig.~\ref{Fig:Fsymbol}
in more detail. Let us specify the five possible labelings of 
three neighboring fusion chain labels $x_{i-1}, x_i, x_{i+1}$,
where in Fig.~\ref{Fig:Fsymbol} we depicted the case where
the site label is $i=2$,
\begin{multline*}
\ket{x_{i-1},x_{i},x_{i+1}} \in \\
\{ \ket{\id,\btau,\id}, \ket{\id,\btau,\btau}, \ket{\btau,\btau,\id},\ket{\btau,\id,\btau},\ket{\btau,\btau,\btau}\} \ .
\end{multline*}
After performing the basis transformation shown in Fig.~\ref{Fig:Fsymbol}, the following labels satisfy 
the fusion rules at each vertex and thus form the new basis
\begin{multline*}
\ket{x_{i-1},\tilde{x}_{i},x_{i+1}} \in \\
\{ \ket{\id,\id,\id}, \ket{\id,\btau,\btau}, \ket{\btau,\btau,\id},\ket{\btau,\id,\btau},\ket{\btau,\btau,\btau}\} \ ,
\end{multline*}
where $\tilde{x}_i$ is the fusion channel of the two neighboring $\btau$ particles.
In the transformed basis, we can project onto the trivial channel, by means of a
projection $P_{i,\id}$, where the subscript $i$ denotes that we are acting on anyons
$i$ and $i+1$, while the label $\id$ denotes we are projecting onto the $\id$ channel.
So, the part of the Hamiltonian acting on anyons $i$ and $i+1$, which we denote by
$H^{i}$, acts on the Hilbert space as
\begin{equation}
\label{Eq:hamiltonian}
\begin{split}
&H^i \ket{x_{i-1},x_{i},x_{i+1}} = \\&
-\sum_{x_i' = \id,\btau}
\left(F^{x_{i-1},\btau,\btau}_{x_{i+1}}\right)^{x_{i}}_{\id}
\left(F^{x_{i-1},\btau,\btau}_{x_{i+1}}\right)^{\id}_{x'_{i}}
\ket{x_{i-1},x'_{i},x_{i+1}} \ .
\end{split}
\end{equation}
Here, we have used that for the su(2)$_k$ anyonic theories we are considering, the
$F$-symbols are their own inverses. Moreover, we projected onto the $\id$
channel, which we favored, because of the overall minus sign.
The total Hamiltonian then simply becomes the sum of \eqref{Eq:hamiltonian} over all
positions
\begin{equation}
\label{Eq:ham}
H = \sum_{i=1}^{L} H^{i} \ ,
\end{equation}
where we assume periodic boundary conditions, i.e. $x_{L} = x_{0}$.

To describe the Hamiltonian of the various types of anyon chains we consider in
this paper, we only have to specify the explicit form of the $F$-symbols (apart from
the fusion rules, which determine the Hilbert space). The explicit
form of the Hamiltonian then follows from Equation \eqref{Eq:hamiltonian}.

\subsection{Galois conjugation and non-unitary models}
\label{Sec:YL-Models}

Now that we have expressed the Hamiltonian in terms of the $F$-symbols, we should
explain how to obtain the $F$-symbols for a given anyon theory. As stated, the $F$-symbols 
transform between two different fusion bases as illustrated in Fig.~\ref{Fig:Fgeneral}. 
As such, the exact form of the these symbols  can be determined self-consistently by 
identifying a circular sequence of basis transformations, 
which yield a set of 
strongly overconstrained nonlinear equations called the `pentagon equations' 
(for a more  detailed exposition see, for instance, Refs.~\onlinecite{Trebst08b} and \onlinecite{as10}).
While finding a solution to these pentagon equations 
is in general a highly non-trivial task, 
it has been shown that 
they allow only  for a {\sl finite} set of inequivalent solutions,
a property which goes under the name of `Ocneanu rigidity', see for instance
Ref.~\onlinecite{eno05}.
For the su(2)$_k$ anyonic theories of interest here, the complete set of possible 
$F$-symbols can be found, e.g., in Ref.~\onlinecite{kr88} where they were obtained
by using quantum group techniques.

The different $F$-symbols are found to have a general form that depends on a single, 
so-called `deformation parameter' $q$ only. 
This deformation parameter  has to be chosen appropriately \cite{kr88}
and  it  turns out  that for the su(2)$_k$ anyonic theory it must be one of the  $(k+2)^{\rm nd}$  
primitive roots of unity, i.e. of the form
\begin{equation}
\label{DefRootOfUnity}
   q = e^{\, p \, \cdot \, 2\pi i/(k+2)} \,,
\end{equation}
where the integer index $p$ runs from $1 \leq p \leq (k+2)/2$ (and $p$ and $k+2$ are relative prime). 
The process of increasing the
index $p$ by one, i.e. going from one root of unity to the next, is what is usually referred to 
as {\sl Galois conjugation}.
 \begin{figure}[t]
\begin{center}
  \includegraphics[width=.9\columnwidth]{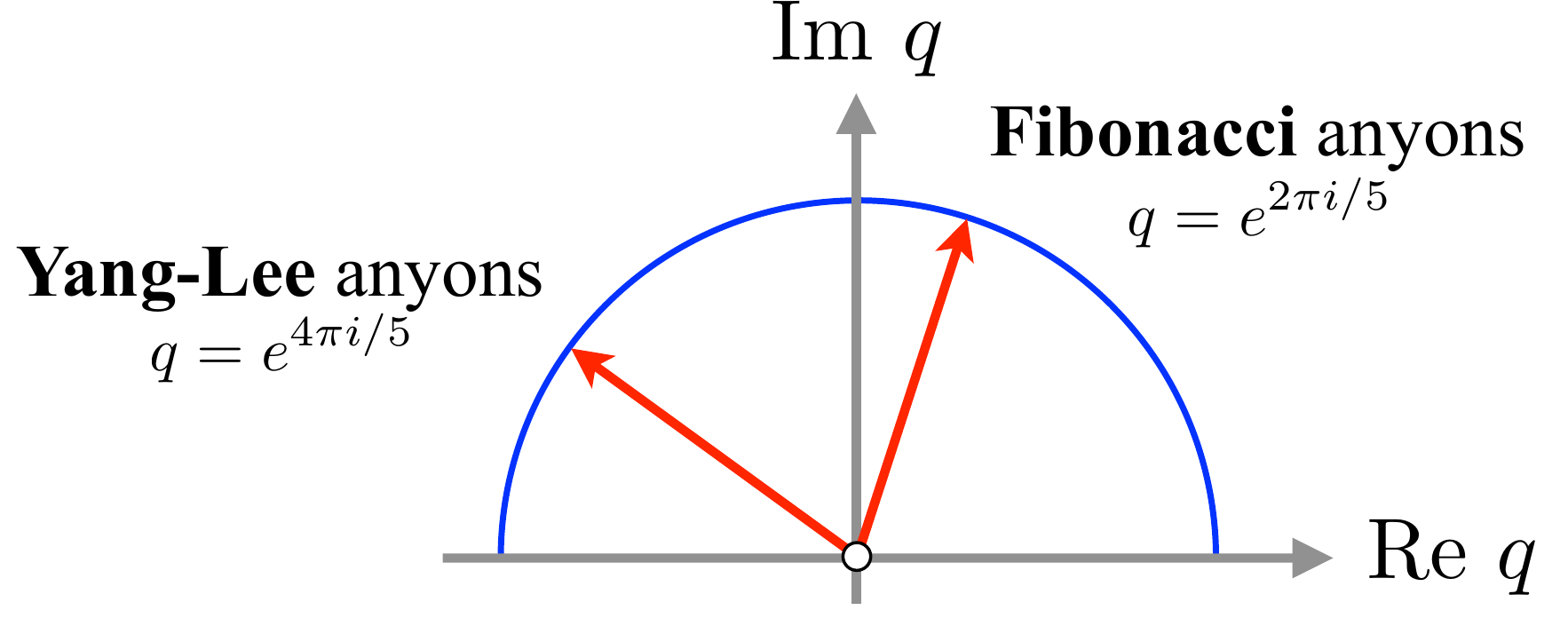}
\end{center}
\caption{The $q$-deformation parameters of Fibonacci and Yang-Lee anyons correspond 
              to different primitive roots of unity.}
\label{Fig:YangLeeAnyons}
\end{figure}
For our example theory, su(2)$_3$, we can thus identify two possible values for $q$, which
are illustrated in Fig.~\ref{Fig:YangLeeAnyons}.
These two Galois conjugated theories corresponding to deformation parameters, $q=e^{2 \pi i /5}$ 
and $q=e^{4 \pi i /5}$, then precisely correspond to the cases of Fibonacci and Yang-Lee anyons, respectively.
The explicit form of the $F$-symbols and in particular the non-diagonal $2\times2$ matrix 
for $F^{\btau,\btau,\btau}_{\btau}$ can then be written \cite{kr88}
in terms of this deformation parameter $q$ as
\begin{equation}
\label{Eq:Fmatgen}
F^{\btau,\btau,\btau}_{\btau} =
\begin{pmatrix}
\frac{1}{q^{-1}+1+q} & \frac{1}{\sqrt{q^{-1}+1+q}} \\
\frac{1}{\sqrt{q^{-1}+1+q}} & \frac{q^{-1}-1+q}{q^{-1}+q}\\
\end{pmatrix} \,.
\end{equation}
For Fibonacci anyons we set $q=e^{2 \pi i /5}$, in which case
$q^{-1}+1+q = 1+2 \cos(2 \pi/5) = (1+\sqrt{5})/2 = \phi$ 
is the golden ratio,
and the $F$-symbol becomes the {\sl unitary} matrix
\begin{equation}
\label{Eq:fsym}
   F_{\rm Fibonacci} 
         =  \begin{pmatrix}   \phi^{-1} & {\phi}^{-1/2} \\
                                     {\phi}^{-1/2} & -\phi^{-1}
          \end{pmatrix} \,.
\end{equation}
The golden ratio, of course, is one solution of the equation
$x^2 = 1+x$, which is an algebraic analog of the fusion rule
$\btau \times \btau = {\bf 1} + \btau$ of the su(2)$_3$ anyonic theory.
The process of taking the Galois conjugate of the original Fibonacci anyon model 
corresponds then simply  to the substitution $\phi \rightarrow -1/\phi$, where $-1/\phi$ 
is the other solution to the equation $x^2 = 1 + x$. 
In terms of the deformation parameter $q$, this amounts to choosing the other possible value of 
$q=e^{4 \pi i /5}$, which indeed yields $q^{-1}+1+ q^{1} = 1+ 2 \cos (4\pi/5) = -1/\phi$. 
The $F$-symbol for Yang-Lee anyons thus becomes the (invertible) 
{\sl non-unitary} matrix
\begin{equation}
\label{Eq:fsymnu}
   F_{\rm Yang-Lee} 
        =  \begin{pmatrix}   -\phi & -i {\phi}^{1/2} \\
                                     -i {\phi}^{1/2} & \phi
          \end{pmatrix} \ .           
\end{equation}

Having obtained the $F$-symbols in both the unitary as well as the non-unitary case,
we can now write down the Hamiltonians for the Fibonacci and Yang-Lee chains.
On the states
$\ket{x_{i-1},x_{i},x_{i+1}} \in \{\ket{{\bf 1},\btau,\bf 1}, \ket{{\bf 1},\btau,\btau},\ket{\btau,\btau,{\bf 1}}\}$
both Hamiltonians act in the same diagonal way, $H^{i} = {\rm diag}\{-1,0,0\}$.
Acting on the states 
$\ket{x_{i-1},x_{i},x_{i+1}} \in \{\ket{\btau,{\bf 1},\btau},\ket{\btau,\btau,\btau}\}$, the Hamiltonians take
the following forms%
\cite{FootnoteGauge}
\begin{eqnarray}
  H^i_{\rm Fibonacci} & = & - \begin{pmatrix} \phi^{-2} & {\phi}^{-3/2} \\
                                     {\phi}^{-3/2} & \phi^{-1}
          \end{pmatrix} \,,
  \nonumber \\
  H^i_{\rm Yang-Lee} & =  & - \begin{pmatrix} \phi^{2} & i{\phi}^{3/2} \\
                                     i{\phi}^{3/2} & -\phi
          \end{pmatrix} \,.
  \label{Eq:HMatrix}
\end{eqnarray}

Before discussing these anyonic models in further detail, we note that while
Galois conjugation changes some aspects of these models, i.e. 
the parameters in their respective Hamiltonians 
get `Galois conjugated', this turns out to be a
rather mild change, since the underlying algebraic structure of 
these models remains largely untouched. 
As a consequence, the non-unitary Yang-Lee chains allow for
an analytic solution similar to their unitary counterparts 
as first obtained for 
the `golden chain' model in Ref.~\onlinecite{Feiguin07}. We will discuss
the details of this analytical solution and its resulting gapless theories in the
next Section.


\subsection{Algebraic structure and analytical solution}
\label{Sec:YL-AnalyticalSolution}

In the original golden chain paper \cite{Feiguin07}, it was shown that the
Hamiltonian \eqref{Eq:ham} 
based on the {\it unitary} $F$-symbols \eqref{Eq:fsym}, 
can be exactly solved. 
Following a similar sequence of steps as in the unitary case
we show here that all non-unitary models allow for an
exact analytical solution as well. We briefly outline these steps in the following,
for a more detailed discussion see the references \onlinecite{Feiguin07},\onlinecite{Trebst08b}.

As a first step, it was noted that
the operators $H^{i}$, i.e. the summands  of the Hamiltonian, form
(upon suitable normalization)
a known representation~\cite{PasquierNPB1987} 
of the Temperley-Lieb algebra~\cite{TemperleyLieb}
with ``d-isotopy"- parameter $d$, 
namely ${\bf e}_i = -d H^{i}$, where $d=\phi$. These operators satisfy the
Temperley-Lieb algeba
\begin{align}
{\bf e}_i^2 &=d \  {\bf e}_i &
{\bf e}_i  {\bf e}_{i\pm1} {\bf  e}_i &= {\bf  e}_i \\
[{\bf e}_i, {\bf e}_j] &=0 \quad {\rm for} \ |i-j|\geq 2 \nonumber
\label{TemperleyLiebRelations}
\end{align}

\begin{figure}[t]
\begin{center}
 \includegraphics[width=.75\columnwidth]{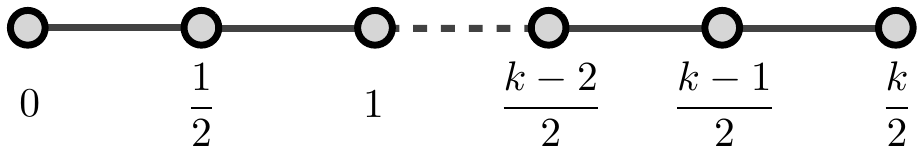}
\end{center}
\caption{The allowed `height' configurations of the chain models. This diagram
is the Dynkin diagram of the Lie algebra $A_k=su(k+1)$. The Fibonacci and Yang-Lee
anyons correspond to $k=3$.}
\label{Fig:AkDiagram}
\end{figure}

To show that this is the case, we will first map the Fibonacci and Yang-Lee
chains onto so-called restricted solid-on-solid (RSOS) or `height' models
\cite{abf84}. To do this, 
we first note 
that one can relate the Fibonacci anyons to the anyonic theory su(2)$_3$, a
theory with four particles (see section \ref{Sec:su2k} for the more general case of
su(2)$_k$), which can be 
labeled by their 'spin' $j=0,1/2,1,3/2$.
The fusion rules of these particles are given in table \ref{Tab:su32fusionrules}.
\begin{table}[ht]
\begin{center}
\begin{tabular}{r|cccc}
$\times$ & $0$ & $1/2$ & $1$ & $3/2$ \\
\hline
$0$ & $0$ & $1/2$ & $1$ & $3/2$\\
$1/2$ & & $0+1$ & $1/2+3/2$ & $1$\\
$1$ & & & $0+1$ & $1/2$\\
$3/2$ & & & & $0$\\
\end{tabular}
\end{center}
\caption{Fusion rules of the su(2)$_3$ theory.}
\label{Tab:su32fusionrules}
\end{table}
One immediately notices that the particle with $l=1$ has the same fusion
rules as the Fibonacci anyon. Indeed, the Fibonacci anyon model can be viewed
as the `integer subset' of the su(2)$_3$ theory. In addition, one notices that
fusing a particle $j$ with the particle $3/2$, 
one finds $j\times 3/2 = 3/2-j$, 
swapping integer `spin' to half integer `spin'. 
This allows one to map the Fibonacci chain,
consisting of `spin'-1 particles, to a chain of `spin'-1/2 particles, by fusing the
labels $x_{2i-1}$ of the odd sites with the particle $3/2$. In this way, the new labels
are constant labelings of a chain consisting of `spin'-1/2 particles.
Performing this map is
advantageous, because the transformed Fibonacci-chain can now directly be
mapped onto a height model, in which the allowed heights take the values
$0,1/2,1,3/2$, which can be seen as the nodes of the Dynkin-diagram $A_4$,
which is given in Fig.~\ref{Fig:AkDiagram}.

The well-known Jones representation of the Temperley-Lieb algebra,
acting on the Hilbert space of the (transformed) Fibonacci chain,
can now be obtained by using the so-called modular S-matrix.\cite{PasquierNPB1987}
Explicitly, this representation is known to take  on the following form:
\begin{equation}
\begin{split}
&{\bf e}_i \ket{x_{i-1},x_{i},x_{i+1}} = 
\sum_{x'_i}  \left((e_i)^{x_{i+1}}_{x_{i-1}}\right)^{x'_i}_{x_i}  \ket{x'_{i-1},x'_{i},x'_{i+1}} \ , \\&
\left((e_i)^{x_{i+1}}_{x_{i-1}}\right)^{x'_i}_{x_i} =
\delta_{x_{i-1},x'_{i-1}}
\delta_{x_{i+1},x'_{i+1}}
\frac{\sqrt{S_{0,x_{i}}}\sqrt{S_{0,x'_{i}}}}{\sqrt{S_{0,x_{i-1}}}\sqrt{S_{0,x_{i+1}}}}
\end{split}
\label{TLrep}
\end{equation}
where $S_{j_1,j_2}$ is the modular $S$ matrix of the su(2)$_3$ theory, 
and the labels $j_1,j_2$ are the `spins' of the corresponding particles, 
which in general take the values
$0,1/2,1,\ldots,k/2$, and in terms of these, the $S$-matrix elements read
\begin{equation}
S_{j_1,j_2} = \sqrt{\frac{2}{k+2}} \sin \left( \frac{(2j_1+1)(2j_2+1)\pi}{k+2} \right) \ . 
\end{equation}

In equation \eqref{TLrep}, the row-label $j_1$ of the $S$-matrix elements 
which appears here takes on
the value $j_1=0$, which is the correct value in case of the (unitary)
Fibonacci chain. In taking the
Galois conjugate, one has to replace this row-label $j_1=0$ by the label $j_1=1$ 
(corresponding to a $\btau$ particle [see also section \ref{Sec:su2k}]).  

The action of the $e_i$'s on the height states, precisely corresponds to the action of the
part of the Hamiltonian acting on the ket $\ket{x_{i-1},x_{i},x_{i+1}}$, which shows that
the operators of Hamiltonian of the Fibonacci chain do indeed form a representation of
the Temperley-Lieb algebra.

We will now employ this observation, and map the Fibonacci chain to an integrable,
two-dimensional statistical mechanics model, namely an RSOS model,
which in the present case is based on the Dynkin diagram $A_4$ -- see
Fig.  \ref{Fig:AkDiagram}. 
In an RSOS model, the degrees of freedom are the heights (or nodes of the
Dynkin diagram). These heights live on the vertices
of the lattice, with the constraint that heights of neighboring
vertices correspond to nodes of the graph which are linked.

\begin{figure}[h]
\begin{center}
  \includegraphics[width=.95\columnwidth]{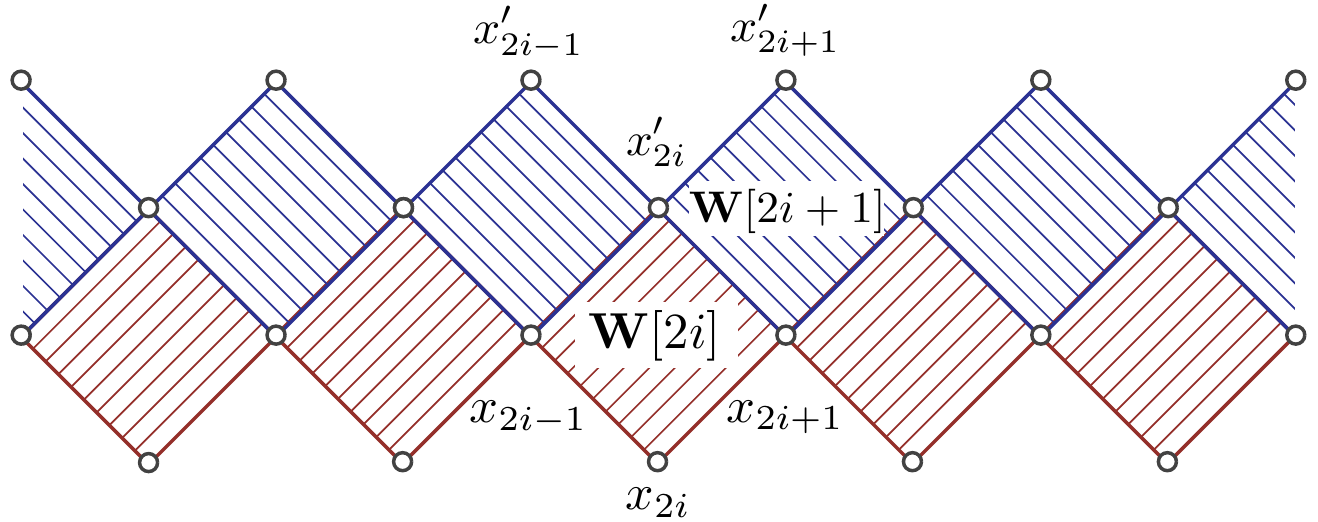}
\caption{The two-row transfer matrix of the RSOS models.}
\label{Fig:tworowtransfermatrix}
\end{center}
\end{figure}

The two-row transfer matrix of the RSOS-model (depicted in
figure \ref{Fig:tworowtransfermatrix}) can be written in terms of the
plaquette weights of the square lattice, namely
${\bf T} = {\bf T}_2 {\bf T}_1$, with 
\begin{align}
{\bf T}_1 &= \prod_{i} {\bf W}[2i] &
{\bf T}_2 &= \prod_{i} {\bf W}[2i+1] \ ,
\end{align}
where the plaquette weights are of the form
\begin{align}
{\bf W}[i]^{\vec{x}'}_{\vec{x}} &=
\left(
\frac{\sin(\frac{p\pi}{k+2}-u)}{\sin(\frac{p\pi}{k+2})} {\bf 1}^{\vec{x}'}_{\vec{x}} +
\frac{\sin(u)}{\sin(\frac{p\pi}{k+2})} {\bf e}[i]^{\vec{x}'}_{\vec{x}} 
\right) 
\label{Eq:TransferMatrix1}
\\
{\bf e}[i]^{\vec{x}'}_{\vec{x}} &= \left( \prod_{j\neq i} \delta_{x'_j,x_j}\right)
\left((e_i)^{x_{i+1}}_{x_{i-1}}\right)^{x'_i}_{x_i}
\label{Eq:TransferMatrix2}
\end{align}
with ${\bf 1}^{\vec{x}'}_{\vec{x}} = \left( \prod_{j} \delta_{x'_j,x_j}\right)$
the identity operator.

The new feature of the {\sl non-unitary} models consists in the fact that 
equations \eqref{Eq:TransferMatrix1} and \eqref{Eq:TransferMatrix2}
for the plaquette weight in the transfer matrix now involve a general
integer $p=2j_1+1$ labeling the primitive root of unity \eqref{DefRootOfUnity}
beyond the unitary case of $p=1$ (corresponding to $j_1=0$) studied earlier.
For a general value of $p$, the
Hamiltonian associated with this lattice model can be obtained by taking the
extreme anisotropic limit, 
namely  $u\rightarrow 0^+$  (see for instance \onlinecite{book:baxter82}). 
One has,
\begin{equation}
{\bf T} = \exp\left(-a({\bf H}+c_1)+O(a^2)\right) \ ,
\end{equation}
with $a = \frac{u}{d \sin(\frac{p\pi}{k+2})}\ll 1$ and $c_1$ an unimportant constant,
which indeed gives that the Hamiltonian is of the form
$H = -\frac{1}{d} \sum_i e_i$, establishing the mapping from the chain to the RSOS models.

We can now use the known results about the phases of the RSOS models, to find the
behavior of the chain models. In the case of the Fibonacci anyons (i.e., when $k=3$
and $p=1$, which appear in the plaquette weights ${\bf W}$ and the parameter $a$),
the anti-ferromagnetic golden chain Hamiltonian can be obtained by taking the limit
$u\rightarrow 0$ with $u$ positive. 
The corresponding (exactly integrable) 
RSOS model\cite{abf84} is critical 
and known\cite{abf84,h84} to be in the universality class of the tri-critical Ising model
described by the minimal model $\mathcal{M}(4,5)$ of conformal field theory.
For the ferromagnetic golden chain, which is obtained from the RSOS
models with $0<|u|\ll 1$ and $u$ negative, 
the critical behavior then turns out \cite{abf84,h84} to be in the universality
class of the $Z_3$ parafermions corresponding to the minimal model
variant
\cite{FootnoteVariant}
$\widetilde{\mathcal{M}}(5,6)$.

In the case of the Galois-conjugated Yang-Lee anyonic chains (corresponding to $k=3$
and $p=2$), the same sequence of steps as above results in a mapping to another family of exactly integrable
RSOS models \cite{fb84} and in this case the gapless theories turn out be
{\sl non-unitary} minimal models \cite{Riggs,Nakanishi}.
In this non-unitary case, the Hamiltonian can be obtained from the two-row transfer
matrices, but in this case, both the anti-ferromagnetic as well as the ferromagnetic
chains are obtained in the limit $u\rightarrow 0$ with positive $u$. The difference in
the Hamiltonians stems from the sign-changes in the isotopy parameter $d$, namely
from positive (but not necessarily bigger than one) in the anti-ferromagnetic case, to
negative in the ferromagnetic case. The critical behavior in these two cases is described
by the (non-unitary) minimal models $\mathcal{M}(3,5)$ and $\mathcal{M}(2,5)$
respectively.


\subsection{Numerical results}
\label{Sec:YL-Numerics}

We have numerically studied the excitation spectra of the Yang-Lee chains by exact diagonalization of 
systems with up to $L=32$ anyons, typically using periodic boundary conditions. 
These excitation spectra not only allow for an independent identification 
of the conformal field theory describing the gapless collective state, as discussed in the previous section, but also reveal
further details about the correspondence between continuous fields and microscopic observables. 
In particular, the low-energy states of a conformally invariant system can be identified with conformal fields 
and the excitation spectrum is expected to take the form
\begin{equation}
  E = E_1 L + \frac{2 \pi v}{L} \cdot \left( -\frac{c}{12} + h + \bar{h} \right) \,,
  \label{Eq:ConformalSpectrum}
\end{equation}
where $h$ and $\bar{h}$ are the (holomorphic and anti-holomorphic) conformal weights of a given CFT with central charge $c$. $E_1$ is a non-universal number, $v$ a non-universal scale factor, and $L$ the length of the chain. 
To match the excitation spectra of the Yang-Lee chains to these CFT predictions we consider the family of so-called minimal models $M(p,p')$ (where $p$ and $p'$ are mutually prime) with central charge
\[
   c = 1 -\frac{6(p-p')^2}{pp'} \,,
\]
and conformal weights
\begin{equation}
   h(r,s) = \frac{(rp-sp')^2 - (p-p')^2}{4pp'} \,,
\end{equation}
where the indices $r$ and $s$ are limited to $1 \leq r < p'$ and $1 \leq s < p$. We note that
the labels $(r,s)$ and $(p'-r,p-s)$ correspond to the same field.

\begin{figure}[t]
\begin{center}
  \includegraphics[width=\columnwidth]{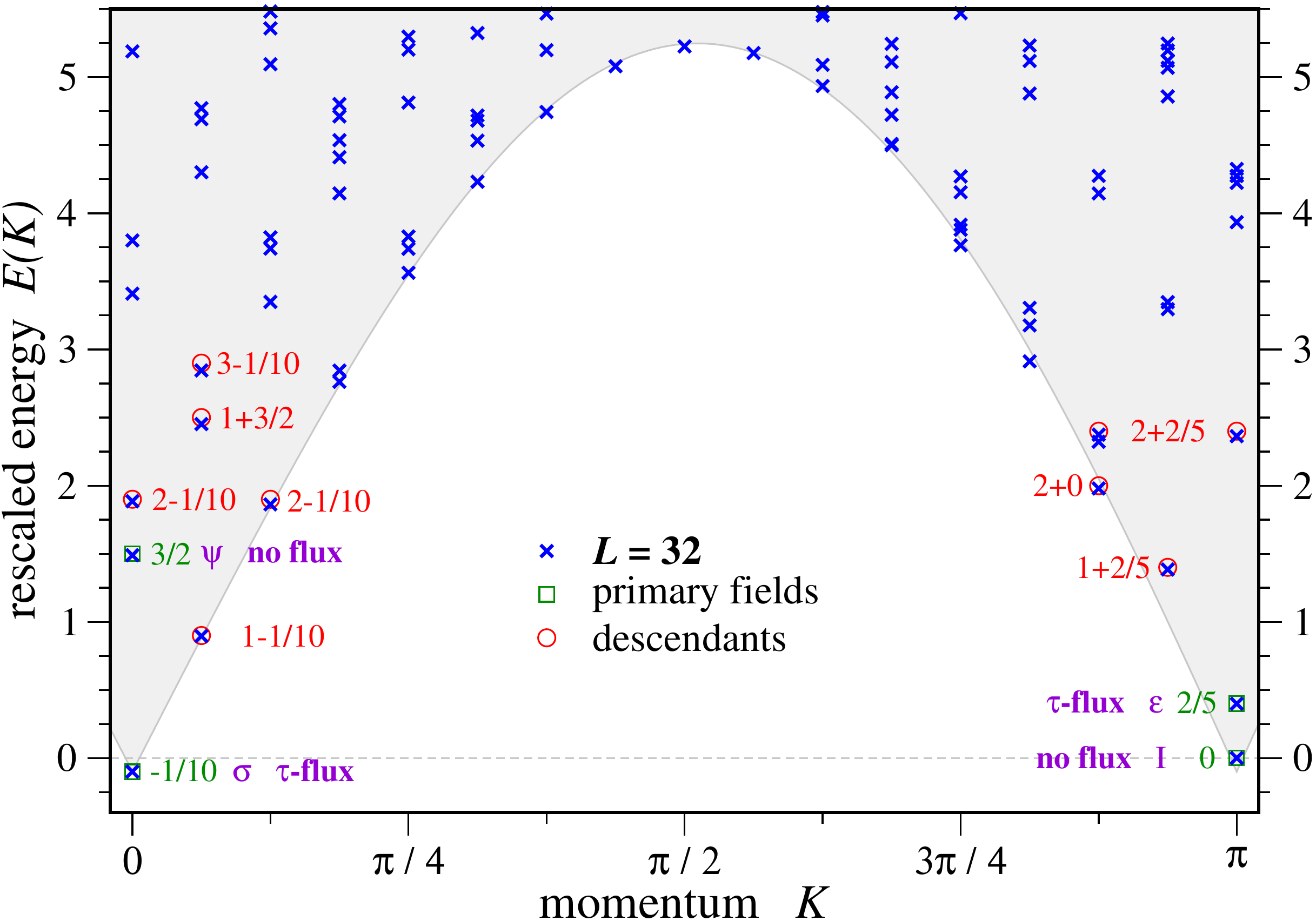}
\end{center}
\caption{
   Conformal excitation spectrum of the `antiferromagnetic' Yang-Lee chain.
   The spectrum matches the non-unitary minimal model $M(3,5)$ 
   with central charge $c=-3/5$, which is often referred to as `Gaffnian' theory.
   Primary fields $I, \sigma, \epsilon, \psi$ of this conformal field theory are indicated by squares, 
   descendant fields by circles.
   We also indicate the `topological flux' of each energy eigenstate, which indicates the topological symmetry sector.
 }
\label{Fig:AFM-Chain}
\end{figure}

\begin{figure}[t]
\begin{center}
  \includegraphics[width=\columnwidth]{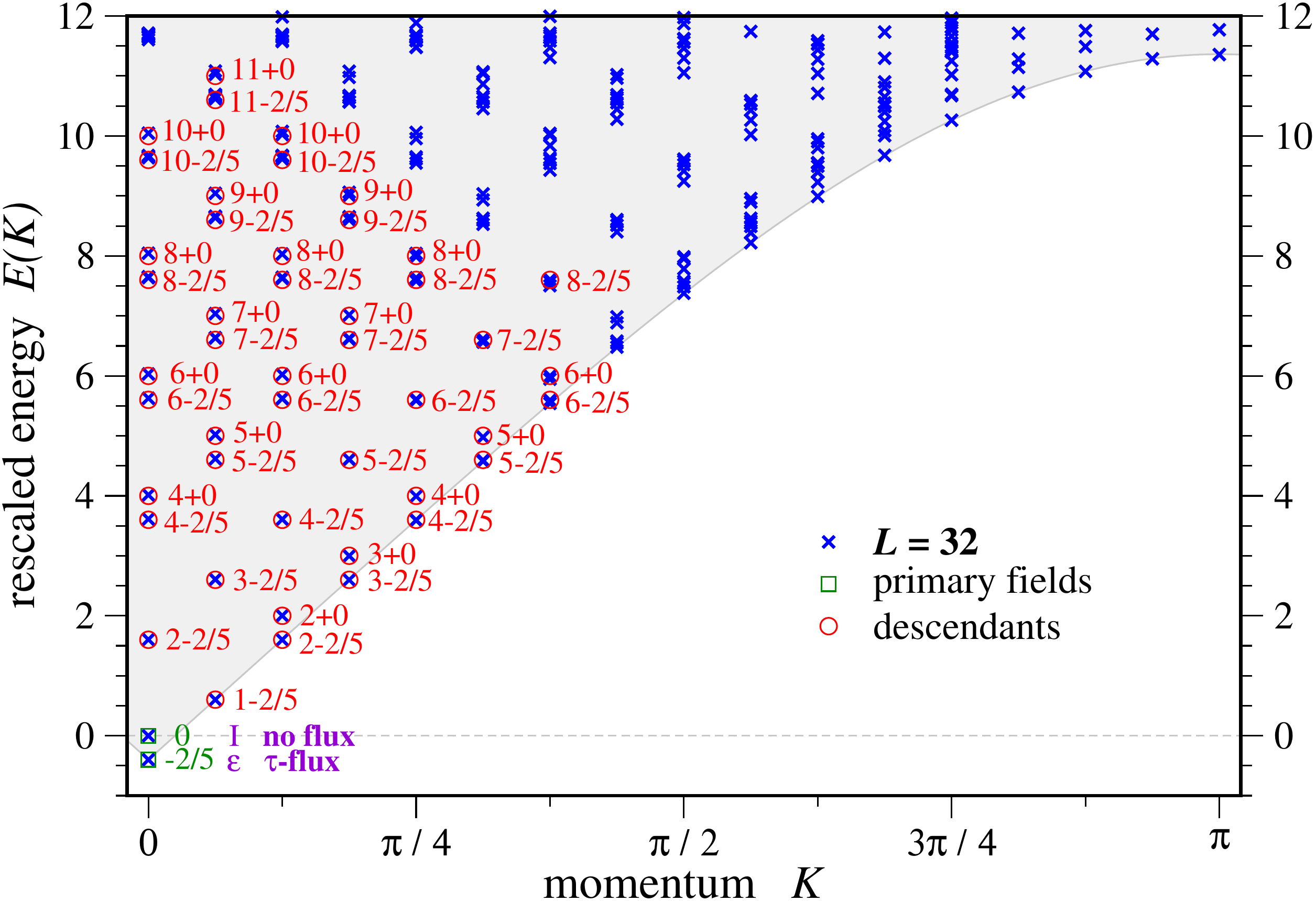}
\end{center}
\caption{
   Conformal excitation spectrum of the `ferromagnetic' Yang-Lee chain.
   The spectrum matches the non-unitary minimal model $M(2,5)$ 
   with central charge $c=-22/5$, which is commonly referred to as `Yang-Lee' theory.
   The primary fields $I, \epsilon$ of this conformal field theory are indicated by squares, 
   descendant fields by circles.
   We also indicate the `topological flux' of each energy eigenstate, which indicates the topological symmetry sector.
 }
\label{Fig:FM-Chain}
\end{figure}

In the following, we will discuss our numerically obtained excitation spectra for `antiferromagnetic' and `ferromagnetic' couplings, which are plotted in Figs.~\ref{Fig:AFM-Chain} and \ref{Fig:FM-Chain}, respectively.


\paragraph*{The antiferromagnetic chain.--}

We first turn to the `antiferromagnetic' chain, for which the pairwise anyon-anyon interaction energetically favors the trivial fusion channel
\[
    \btau \times \btau \to {\bf 1} \,.
\]
The conformal field theory describing the critical behavior of this model is
the non-unitary minimal model $\mathcal{M}(3,5)$
with central charge $c = -3/5$, which is also referred to as the `Gaffnian' theory \cite{Gaffnian}.
The four primary fields of this CFT and their respective scaling dimensions
$\Delta = h+\bar{h}$ are
\begin{equation}
\begin{tabular}{c|cccc}
& $\sigma$ & $I$ & $\epsilon$ &  $\psi$ \\
\hline
$\Delta$ &   -1/10 & 0 & 2/5 &  3/2
\end{tabular}
\end{equation}
with the non-trivial fusion rules
\begin{align*}
\sigma \times \sigma  &=  I + \epsilon &
\sigma \times \epsilon  &=  \sigma + \psi &
\sigma \times \psi &= \epsilon \\
& & \epsilon \times \epsilon  &=  I + \epsilon & 
\epsilon \times \psi &= \sigma \\
& & & & \psi \times \psi  &=  I 
\end{align*}
For completeness, we give the conformal dimensions of the fields
(with minimal model labeling) in table \ref{Tab:KacM3525}, and note
that this model is a particular Galois conjugate of the su(2)$_3$ CFT.

To identify the gapless theory numerically, we typically perform the following
procedure: We first look at the two lowest energy eigenvalues in the spectrum, 
$E_0$ and $E_1$, and by identifying the energy gap $\Delta E = E_1 - E_0$
with the difference of the two lowest scaling dimensions we can identify the
non-universal scale factor $2\pi v/L$ in \eqref{Eq:ConformalSpectrum}, which we subsequently set 
to 1 thereby rescaling the entire energy spectrum. This identification of the 
two lowest energy eigenvalues with conformal operators also allows to identify
an overall energy shift, e.g. setting the energy of the trivial operator $I$ with
scaling dimension $h(1,1)+\bar{h}(1,1) = 0$ to zero. 
In the case at
hand, there is only one negative scaling dimension, so the lowest energy
corresponds to $-2h_{\rm min}=-1/10$, while the second lowest state corresponds
to the identity operator, with zero energy. At this point, all the energies are fixed,
and indeed the rescaled and shifted numerical spectrum is found to reproduce the 
position of the (other) primary fields (indicated by green squares in Fig.~\ref{Fig:AFM-Chain}), 
as well as the descendants (indicated by red circles in Fig.~\ref{Fig:AFM-Chain}).

\begin{table}[b]
  \begin{tabular}{r|cc}
  \hline \hline
  \multicolumn{3}{c}{$\mathcal{M}(3,5)$} \\
  \hline 
  $h(r,s)$ & $s=1$ & $2$ \\
  \hline
  $r=1$ & 0 & 3/4 \\
  3         & 1/5 & -1/20 \\
  \hline \hline
  \end{tabular}
  \hskip 8mm
  \begin{tabular}{r|c}
  \hline \hline
  \multicolumn{2}{c}{$\mathcal{M}(2,5)$} \\
  \hline 
  $h(r,s)$ & $s=1$ \\
  \hline
  $r=1$ & 0 \\
  3         & -1/5 \\
  \hline \hline
  \end{tabular}
   \caption{Kac table of conformal weights for the non-unitary minimal model $\mathcal{M}(3,5)$ and
   $\mathcal{M}(2,5)$. We only displayed fields with odd $r$ labels, to avoid duplicates.}
  \label{Tab:KacM3525}
\end{table}


\paragraph*{The ferromagnetic chain.--}

We now turn to the `ferromagnetic' chain, for which the pairwise anyon-anyon interaction energetically favors the $\btau$-fusion channel
\[
    \btau \times \btau \to \btau \,.
\]
The critical theory is the non-unitary minimal model $\mathcal{M}(2,5)$
with central charge $c = -22/5$, which is commonly referred to as  the `Yang-Lee' theory \cite{YangLee,Cardy}.
The two primary fields of this CFT and their respective scaling dimensions
$\Delta = h+\bar{h}$ are
\begin{equation}
\begin{tabular}{c|cc}
& $I$ & $\epsilon$ \\
\hline
$\Delta$ & $0$ & $-2/5$ 
\end{tabular}
\end{equation}
with the non-trivial fusion rule
$$\epsilon \times \epsilon = I + \epsilon$$
Again we give the conformal dimensions of the fields
(with minimal model labeling) in table \ref{Tab:KacM3525}
for completeness.

The spectrum of this theory, after the appropriate shift and rescaling
of the energy is displayed in figure \ref{Fig:FM-Chain}, which beautifully
reproduces the primary fields, as well as descendants to a high level,
constituting the spectrum of the Yang-Lee model. 


\subsection{Topological symmetry}
 \label{Sec:TopSym}

Before considering further generalizations of the Yang-Lee chains we mention another peculiarity
of these anyonic chains. Like their unitary counterparts the Yang-Lee chains exhibit an unusual,
non-local symmetry. This symmetry, which was dubbed a `topological symmetry' in the context 
of the golden chain model \cite{Feiguin07}, corresponds to the operation of commuting a $\btau$ 
particle through all particles of the chain. The so-defined operator, which we denote by $Y$, is
found to commute with the Hamiltonian and for the su(2)$_3$ theory has two distinct eigenvalues, 
thus defining two symmetry sectors. Its matrix form is given by 
\begin{equation}
  \langle x_0',\ldots,x_{L-1}'| Y | x_0, \ldots, x_{L-1}\rangle 
  = \prod_{i=0}^{L-1} \left( F_{x'_{i+1}}^{\btau x_i\btau}\right)^{x_i'}_{x_{i+1}} \,.
\end{equation}
This definition solely in terms of the $F$-symbols immediately suggests a generalization of this 
symmetry to the case of the non-unitary Yang-Lee models studied above by simply replacing the
$F$-symbols with their non-unitary counterparts \eqref{Eq:fsymnu}.
For both the unitary and non-unitary variants the two eigenvalues of the respective topological 
symmetry operator are $y_1 = \phi$ and $y_2 = -1/\phi$. In the unitary case these are identified 
as no-flux / $\btau$-flux symmetry sectors. This assignment is simply reversed in the Galois 
conjugated, non-unitary case.

For the unitary models it has been demonstrated that this topological symmetry  {\sl protects} 
the gapless ground state of the interacting anyon chain model \cite{Gils09}:
it was shown that all relevant operators (in a renormalization group sense) that have otherwise identical 
quantum numbers as the ground state, e.g. the same momentum, fall into different topological symmetry 
sectors. 
We have performed a similar symmetry analysis for the Yang-Lee chains at hand.
For chains with either antiferromagnetic or ferromagnetic couplings, we have evaluated the 
eigenvalue of the topological symmetry $Y$ for all eigenvectors of the Hamiltonian and thereby 
assigned topological symmetry sectors to the primary fields of the conformal field theory describing 
their energy spectrum. 
These assignments are given in Figs.~\ref{Fig:AFM-Chain} and \ref{Fig:FM-Chain} for 
antiferromagnetic and ferromagnetic chain couplings, respectively. 
A situation similar to the unitary models emerges: For both signs of the coupling the ground state
with conformal dimension $h + \bar{h} < 0$ is found to be in the topologically non-trivial (or $\btau$-flux) sector, 
while  all other primary fields with the same momentum are found to be in the topologically trivial 
(or no-flux) sector. 

For the unitary anyon chains this topological protection mechanism has subsequently been cast in a broader
physical picture \cite{Gils09} interpreting the gapless modes of the anyonic chains as edge states 
at the spatial interface of two topological liquids, and the conclusion that anyon-anyon interactions
result in a splitting of the topological degeneracy for a set of non-Abelian anyons and the nucleation 
of distinct topological liquid within the parent liquid of which the anyons are quasiparticle
excitations \cite{Gils09,Ludwig10}.
The similarity of our results for the topological symmetry properties of the non-unitary anyon chains
thus raises the question whether a similar interpretation would also hold for the non-unitary systems at 
hand.


\subsection{su(2)$_k$ theories and generalized Galois conjugation}
\label{Sec:su2k}

\begin{table*}[t]
  \begin{tabular}{c|c|c|c}
     \hline
     \hline
     \multicolumn{4}{c}{\bf su(2)$_k$ chains} \\
     \hline
     coupling & CFT & central charge & $d$-isotopy \\
     \hline
     AFM & $\mathcal{M}(k+1,k+2)$ & $c = 1 - \frac{6}{(k+1)(k+2)}$ & $2\cos{\left( \frac{\pi}{k+2} \right)}$ \\
     FM   & $\mathcal{Z}_k$ parafermion & $c=2\frac{k-1}{k+2}$ & $2\cos{\left( \frac{\pi}{k+2} \right)}$ \\
     \hline 
     \multicolumn{4}{c}{\bf Yang-Lee chains}\\
     \hline
     coupling & CFT & central charge & $d$-isotopy \\
     \hline
     AFM & $\mathcal{M}(k,k+2)$ & $c = 1 - \frac{24}{k(k+2)}$ & $2\cos{\left( \frac{2\pi}{k+2} \right)}$ \\
     FM   & $\mathcal{M}(2,k+2)$ & $c = 1 - \frac{3k^2}{k+2}$ & $2\cos{\left( \frac{k\pi}{k+2} \right)}$ \\
     \hline
     \multicolumn{4}{c}{{\bf Generalized Galois conjugates} with $2 \leq p < (k+2)/2$}\\
     \hline
     coupling & CFT & central charge & $d$-isotopy \\
     \hline
     AFM & $\mathcal{M}(k+2-p,k+2)$ & $c = 1 - \frac{6p^2}{(k+2-p)(k+2)}$ & $2\cos{\left( \frac{p\pi}{k+2} \right)}$ \\
     FM   & $\mathcal{M}(p,k+2)$ & $c = 1 - \frac{6(k+2-p)^2}{p(k+2)}$ & $2\cos{\left( \frac{(k+2-p)\pi}{k+2} \right)}$ \\
     \hline
     \hline
  \end{tabular}
  \caption{Overview of the gapless theories for the various chain models.
  }
  \label{Tab:Chainresults}
\end{table*}

While we have focused our discussion in the preceding sections 
on the Fibonacci / Yang-Lee anyonic theories for which $k=3$, 
our results can be readily generalized to a much larger class of anyonic 
theories, so called su(2)$_k$ Chern-Simons theories. 
These theories, which are certain quantum deformations \cite{SU2q} of SU(2), have a {\sl finite}
number of representations which are identified with the (anyonic) particle content of the theory.
For a given (integer) deformation parameter $k$ these representations/particles are labeled by a 
generalized angular momentum
\[  
	j=0,\frac{1}{2},1,\ldots,\frac{k-1}{2},\frac{k}{2} \,.
\] 
The fusion rules for combining two of these momenta (or their associated particles) 
then resemble the tensor product rules of ordinary SU(2) angular momenta, and 
explicitly read
\begin{equation}
	j_1 \times j_2 = \sum_{j_3=|j_1-j_2|}^{\min(j_1+j_2,k-j_1-j_2)} j_3\ .
	\label{Eq:su2kfusion}
\end{equation}
For all $k \leq 2$ the first non-trivial fusion rule (i.e. a fusion rule with multiple
fusion outcomes) is encountered for the particle carrying generalized angular
momentum $j=1/2$, for which one finds
\begin{equation}
  	\frac{1}{2} \times \frac{1}{2} = 0 + 1 \,.  
	\label{Eq:1/2fusion}
\end{equation}
Upon further inspection of the fusion rules (see Table \ref{Tab:su32fusionrules} for an example) 
one notes that  for anyonic theories with odd $k$ one can map all half-integer
representations onto integer representations by fusing the half-integer representations 
with the highest representation of the theory, the so-called `simple current' with
angular momentum $\frac{k}{2}$. 
In particular, this mapping identifies the fusion rules of the particle carrying angular
momentum $j=1/2$ with those of the particle carrying angular momentum $j=(k-1)/2$
and Eq.~\eqref{Eq:1/2fusion} becomes
\begin{equation}
  \frac{k-1}{2} \times \frac{k-1}{2} = 0 + 1 \,.
	\label{Eq:1/2fusion-integer}
\end{equation}
Note that for odd $k$ the fusion rules are now given entirely in terms of integer 
representations \cite{FootnoteIntegerRepresentations}. 
For the su(2)$_{k=3}$ theory this mapping thus turns the fusion rules of the $j=1/2$
particle into those of the $j=1$ particle, i.e. $1 \times 1 = 0 + 1$, which is the
momentum representation of the non-trivial fusion rule in Eq.~\eqref{Eq:FusionRules}. 
One can thus equally identify the $\btau$ particle of the Fibonacci / Yang-Lee theories 
with both the $j=1$ and $j=1/2$ angular momenta in su(2)$_3$. 

The su(2)$_k$ generalizations of the Fibonacci / Yang-Lee chains, which we want
to consider in the following, are chains of particles carrying angular momentum
$j=1/2$, which we can alternatively identify with particles carrying angular
momentum $j=(k-1)/2$. 
The microscopic chain model is then constructed in an identical way to what 
we outlined for the case of the Fibonacci / Yang-Lee theories in Section \ref{Sec:YL-Chains},
i.e. we first build a Hilbert space by considering admissible labelings of the 
fusion chain basis for an incoming $j=1/2$ (or $j=(k-1)/2$) particle and then derive
a Hamiltonian via the local projection onto one of the two fusion outcomes in
Eqs.~\eqref{Eq:1/2fusion} or \eqref{Eq:1/2fusion-integer}, respectively.
Our notion for the sign of the interaction also remains unchanged, with 
`antiferromagnetic' interactions favoring fusion of neighboring particles into 
a total $j=0$ state and `ferromagnetic' interactions favoring fusion into 
a total $j=1$ state. 

\paragraph*{Unitary models.--}
For the unitary case, such su(2)$_k$ generalizations of the golden chain model
were first discussed in Ref.~\onlinecite{Feiguin07} and further explored in 
Ref.~\onlinecite{Gils09}. 
Like their su(2)$_3$ counterpart these generalized anyon chains are found to
be exactly solvable and their gapless collective states can be described in terms
of conformal field theories. 
To be specific, the critical theory for the generalized antiferromagnetic chains 
is known to be the minimal model $\mathcal{M}(k+1,k+2)$ with central charge $c=1-6/((k+1)(k+2))$,
while for ferromagnetic chains it is known to be the $Z_k$ parafermion theory with central charge
$c=2(k-1)/(k+2)$. In both cases, the terms in the Hamiltonian are found to form a
Temperley-Lieb  algebra with  isotopy parameter $d=2\cos(\pi/(k+2))$.

\paragraph*{Non-unitary models.--}

 \begin{figure}[b]
\begin{center}
  \includegraphics[width=.9\columnwidth]{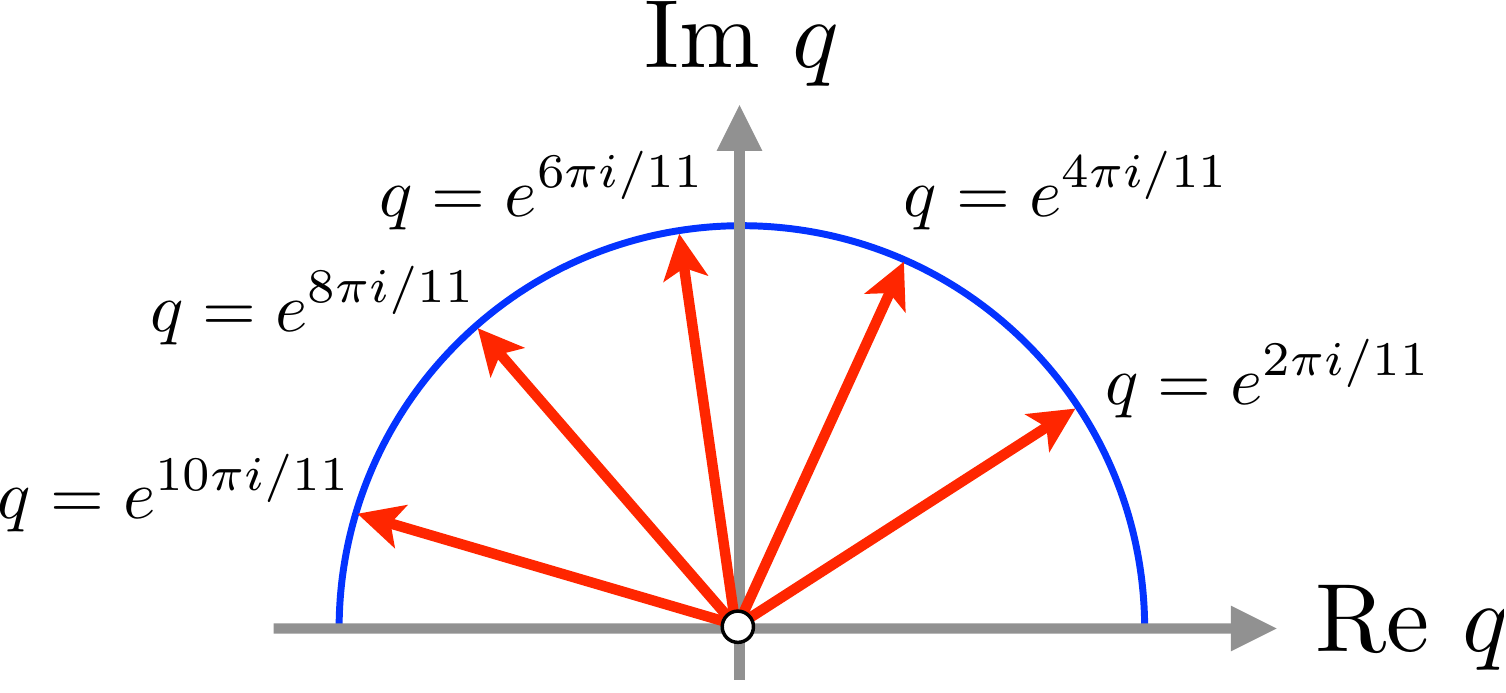}
\end{center}
\caption{The $q$-deformation parameters for a general su(2)$_k$ theory correspond 
              to different primitive roots of unity. Only the first primitive root in this sequence, e.g. 
              $q=e^{2\pi i/11}$ in the illustrated example of $k=9$, gives rise to a unitary model.}
\label{Fig:GeneralizedGaloisConjugation}
\end{figure}

In turning to the Galois conjugated models of these unitary su(2)$_k$ anyon chains, we 
first recall that for a given su(2)$_k$ anyonic theory there are several allowed $q$
deformation parameters 
(see also Sec.~\ref{Sec:YL-Models}) 
\begin{equation}
   q = e^{\, p \, \cdot \, 2\pi i/(k+2)} 
\end{equation}
as illustrated in Fig.~\ref{Fig:GeneralizedGaloisConjugation}.
For each integer index $1 \leq p \leq (k+2)/2$ (and the further restriction that $p$ and $k+2$ 
are relative prime) the 
corresponding $q$ deformation parameter establishes a solution of the pentagon equations 
of the underlying su(2)$_k$ theory and defines the explicit form of the $F$-symbols.
The unitary su(2)$_k$ models discussed above all correspond to the first primitive root, 
i.e. $p=1$, in this sequence. The other primitive roots (with $p \geq 2$) then give rise to the 
Galois conjugated models with non-unitary $F$-symbols and non-Hermitian microscopic 
Hamiltonians constructed using these $F$-symbols. 
In particular, we note that for $k=2$ there is no such Galois conjugated theory, 
for $k=3$ there is exactly one Galois conjugated theory (the Yang-Lee 
theory discussed above), and for $k \geq 4$ there are in general {\sl multiple}
Galois conjugated theories.

Like their unitary counterparts all Galois conjugated su(2)$_k$ chain models 
permit an analytical solution.
Since the line of argument closely follows the one
given in Section~\ref{Sec:YL-AnalyticalSolution} for $k=3$, we will be brief
here, and merely state the main arguments and results.
In order to simplify the presentation we will
further focus our arguments to anyonic theories 
with odd $k$ which allows to restrict the discussion 
to those which possess only integer (not half-integer) values
of ``angular momenta" $j$.
(Even values of $k$ work in an analgous manner.)
The Galois conjugated su(2)$_k$ chain models can then be directly
mapped onto the height representations of certain restricted solid on solid models
where the heights take the values $0,1/2,\ldots,k/2$ in the Dynkin diagram $A_k$
illustrated in Fig.~\ref{Fig:AkDiagram}. The Hamiltonian terms acting in this height representation
then again form a Temperley-Lieb algebra \cite{PasquierNPB1987} with an isotopy
parameter which now turns out to depend on the sign of the interactions.
For antiferromagnetic coupling this generalized isotopy parameter is found to be 
$d = 2 \cos(p \pi/(k+2))$ and the conformal theory describing gapless interacting
system is the non-unitary minimal model $\mathcal{M}(k+2-p,k+2)$ with
central charge~\cite{FootnoteNonUnitary} $c = 1 - \frac{6p^2}{(k+2)(k+2-p)}$.
For ferromagnetic couplings the isotopy parameter becomes 
$d=2\cos{\left(\frac{(k+2-p)\pi}{k+2}\right)}$ and the gapless theory is 
the non-unitary minimal model $\mathcal{M}(p,k+2)$ with 
central charge $c = 1 - \frac{6(k+2-p)^2}{p(k+2)}$. In the case of the Galois-conjugated models,
the central charges are related to the quantum dimensions in the following
way\cite{PasquierNPB1987,df84,n84} 
Writing the quantum dimension as
$d=2\cos (p \pi/ (k+2))= 2\cos (\pi/\delta)$ (where $\delta$ is a fraction for the
Galois-conjugated models), one can obtain the central charge via
$c = 1-\frac{6}{\delta(\delta-1)}$. Note that this is also the correct result for the
anti-ferromagnetic unitary chain with $p=1$, but the central charge for the
ferromagnetic unitary chains is given by $c=2-6/\delta$.

Our results for unitary and non-unitary models are summarized in Table
\ref{Tab:Chainresults}. We note that for $p=2$ and $k=3$ we exactly reproduce 
the results of the Yang-Lee chains discussed above.


\section{Doubled Yang-Lee models}
\label{Sec:YL-Ladders}

\subsection{The ladder Hamiltonian}
\label{Sec:LadderHam}

\begin{figure}[b]
\begin{center}
  \includegraphics[width=.75\columnwidth]{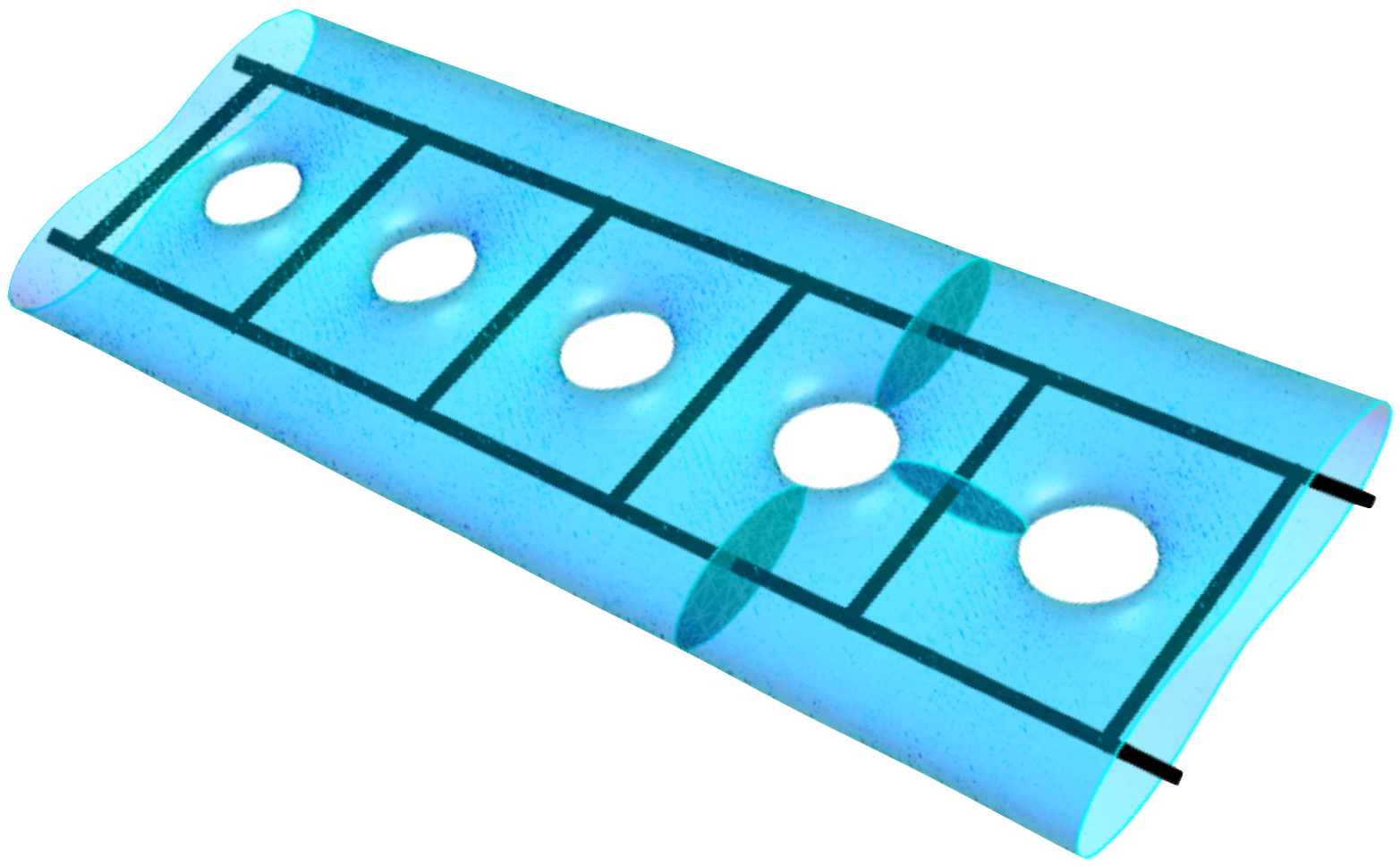}
\end{center}
\caption{
   The geometry of the Fibonacci and Yang-Lee ladders. To discuss the physics,
   it is enlightening to `thicken' the ladder, and consider the two
dimensional surface thus obtained.
}
\label{Fig:YangLeeLadder}
\end{figure}

In this second part of the manuscript, we turn to `quantum double' variants 
of the anyonic chains discussed in the first part. The unitary incarnations of these
quantum double models have been introduced in the context of exotic quantum
phase transitions in time-reversal invariant systems that are driven by topology
fluctuations \cite{Gils09b}. The specific model is defined on a ladder geometry,
shown in figure \ref{Fig:YangLeeLadder}.

The Hamiltonian
\begin{equation}
  H_{\text{ladder}} = -J_r \sum_{\text{rungs } r} \delta_{l(r),\mathbf{1}} - J_p \sum_{\text{plaquettes } p} \delta_{\phi(p),\mathbf{1}}
\end{equation}
consists of two competing terms.

The first term favors the trivial label $\mathbf{1}$ on each rung of the ladder,
while the second term favors the no-flux state for all plaquettes.
As shown in Ref.~\onlinecite{Gils09b}, the projector onto the flux
through a square plaquette can be expressed in terms of the
unitary/non-unitary F-matrices
\eqref{Eq:fsym}/\eqref{Eq:fsymnu}. This term is equivalent to the
plaquette term in the Levin-Wen models\cite{lw05}, which are defined on
a different lattice, the honeycomb lattice. 

Explicitly, the plaquette term reads
\begin{multline}
\delta_{\phi(p),\id}
\Biggl\lvert
\hspace{.1cm}
\raisebox{-.65cm}{\includegraphics[width=2.5cm]{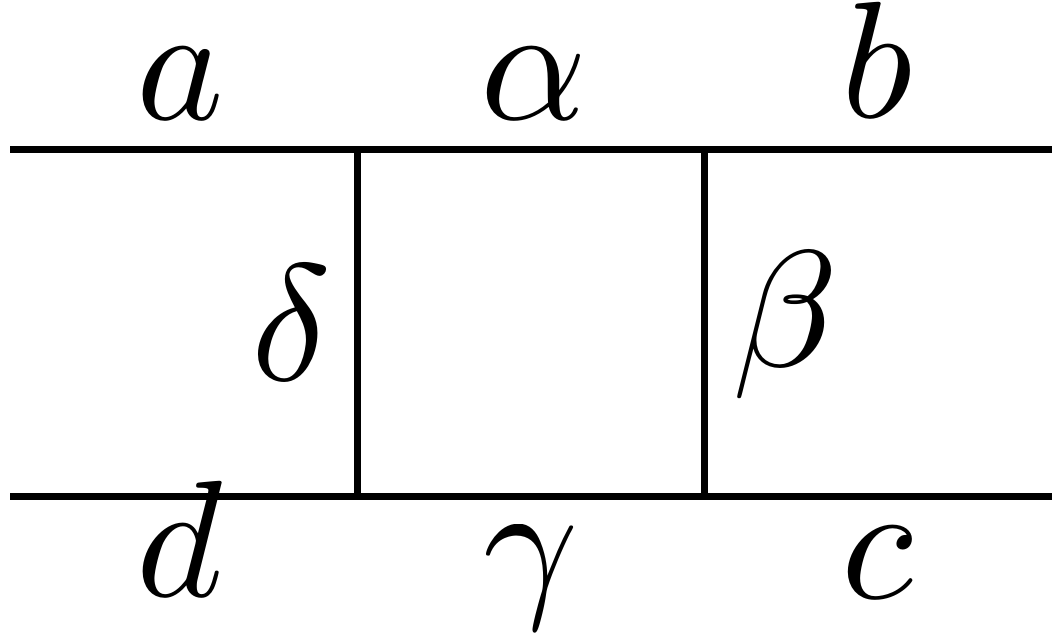}}
\hspace{.1cm}
\Biggr\rangle = \\
\sum_{s=\id,\btau} \frac{d_s}{D^2}\sum_{\alpha',\beta',\gamma',\delta'}
\left( F^{\alpha' s \delta}_{a} \right)_{\alpha}^{\delta'} 
\left( F^{\beta' s \alpha}_{b} \right)_{\beta}^{\alpha'} \\
\times
\left( F^{\gamma' s \beta}_{c} \right)_{\gamma}^{\beta'}
\left( F^{\delta' s \gamma}_{d} \right)_{\delta}^{\gamma'}
\Biggl\lvert
\hspace{.1cm}
\raisebox{-.65cm}{\includegraphics[width=2.5cm]{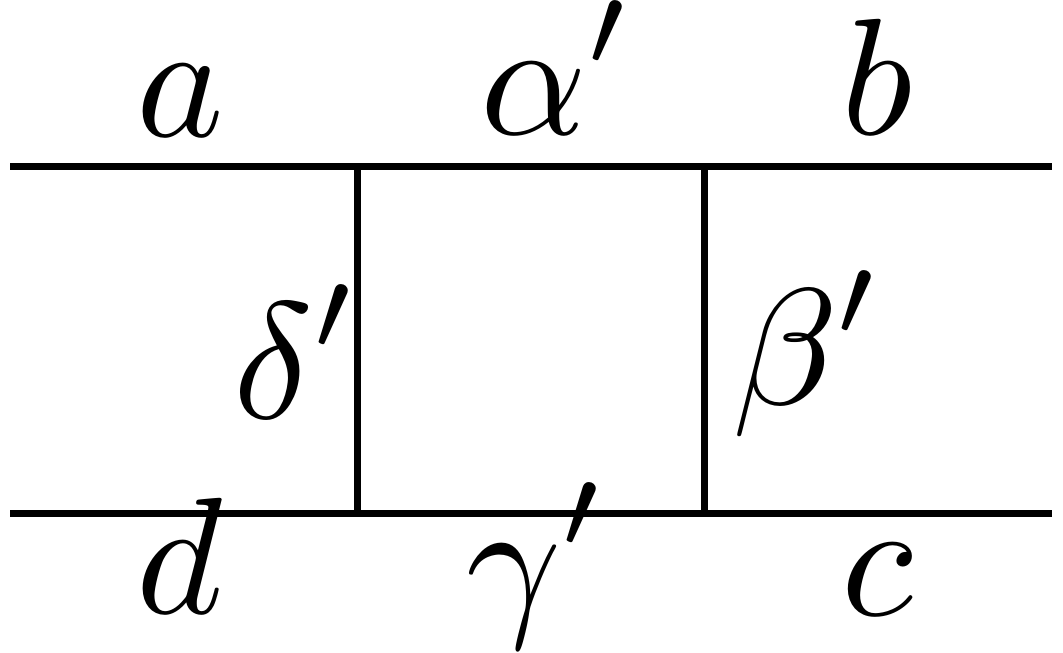}}
\hspace{.1cm}
\Biggr\rangle \ ,
\end{multline}
where $d_s$ denotes the quantum dimension of particle type $s$, i.e.
$d_\id = 1$ and $d_\btau = \phi$. $D$ denotes the total quantum dimension, 
$D = \sqrt{d_\id^2+d_\btau^2} = \sqrt{2+\phi}$ for Fibonacci (as well as Yang-Lee)
anyons. The latin and greek labels denote the degrees of freedom, and any of
these takes one of the 
values $\{ \id,\btau \}$. We note that the Hilbert space of the
ladder models consists of all possible labelings of the rungs and the legs, such that
at each vertex, the Fibonacci fusion rules are obeyed.

With this description of the ladder models, we can easily
go back and forth between the Fibonacci
anyon ladder, and the Yang-Lee anyon ladder, 
simply by choosing the corresonding set of $F$-symbols, 
namely equations \eqref{Eq:fsym} or \eqref{Eq:fsymnu} respectively.


\subsection{The phase diagram}
\label{Sec:PhaseDiag}

\begin{figure}[t]
\begin{center}
  \includegraphics[width=\columnwidth]{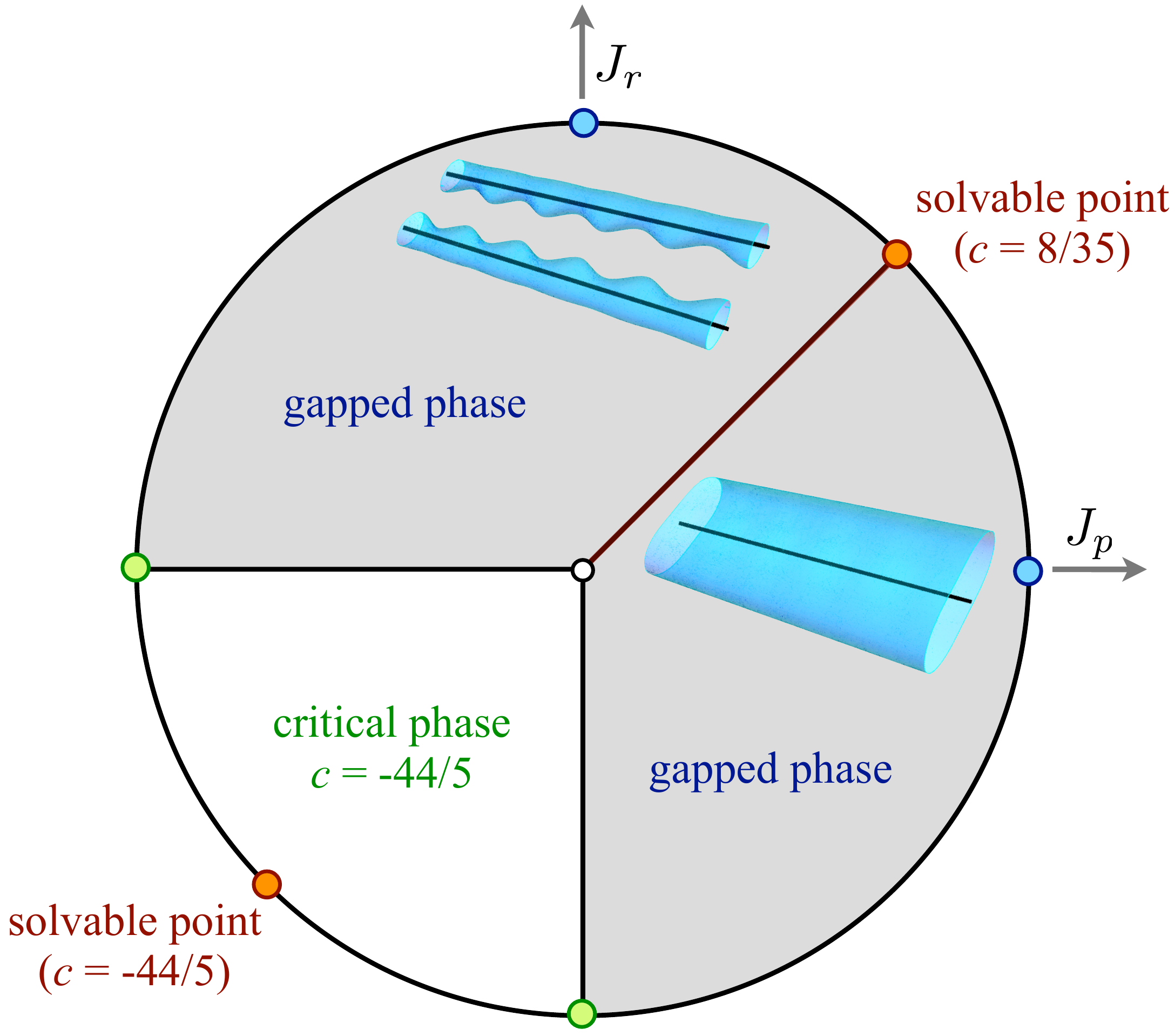}
\end{center}
\caption{
   Phase diagram of the doubled Yang-Lee chain.
 }
\label{Fig:PhaseDiagram-DoubledChain}
\end{figure}

To discuss the phase diagram 
of both the original and Galois conjugated model, shown in figure
\ref{Fig:PhaseDiagram-DoubledChain}, we parametrize the couplings
$J_r=\sin\theta$ and $J_p=\cos\theta$ in terms of an angle $\theta$.

%
%

We start with the first gapped phase, for $\pi/4 < \theta < \pi$, most easily discussed at
the special point $\theta = \pi/2$, or $J_r=1$ and $J_p=0$. In the ground state at this point,
all rungs have the trivial label. In the two-dimensional surface
geometry, this means that no $\btau$-fluxes go through the rungs, implying
that the rungs can be completely pinched off, changing the geometry to
that of two, disconnected cylinders, one for each leg of the ladder. Each of
the cylinders accommodates two ground states, either with or without $\btau$-flux,
leading to a four-fold ground state degeneracy. The lowest (gapped)
excited state consists of configurations where one rung of the ladder
`contains' a $\btau$ flux. 

In the other extended gapped phase, for $-\pi/2<\theta<\pi/4$, let us discuss
the special point $\theta = 0$, or $J_p=1$ and $J_r=0$. Here the situation is
reversed, and no $\btau$-fluxes
go through the {\em plaquettes}. Thus they can be closed of, giving
rise to a geometry consisting of a single cylinder. Again, this cylinder
can accommodate two ground states, with or without flux, resulting in a two-fold
degenerate ground state. The lowest (gapped) excited state consists
of configurations with a single $\btau$-flux going through a plaquette, effectively
piercing a hole through the cylinder.

Precisely at the point where both couplings are equal in strength,
the gap closes, and the system is critical. At this point, the geometry
is fluctuating at all length scales, interpolating between the two extremes
of having one or two cylinders, respectively. In addition, also precisely at
this point, the (critical) model is exactly solvable, as explained in
the next section.

Finally, for $\pi<\theta<3\pi/2$, there is an extended critical region, which is
characterized by another exactly solvable point, at $\theta = 5\pi/4$, where
both couplings are again of equal strength, but negative.


\subsection{Analytical solution}
\label{Sec:CriticalPoints}
The analytic solution of the Fibonacci ladder model at the two exactly solvable points
has been described in detail in
(the supplementary material of) reference \onlinecite{Gils09b}. Thus, we will be
rather brief here, and point out some of the crucial steps of the mapping of the
ladder model onto an exactly solved (restricted solid-on-solid) model. 
We will then quickly mention in which way this solution of the unitary Fibonacci
case is changed, if one takes the Galois conjugate, to go to the case of
the Yang-Lee ladder model. The most important step in the analytic solution of the
model, is to perform a basis transformation on the Hilbert space. After this
transformation, the model can be mapped onto a restricted solid-on-solid model,
where in this case, the allowed height variables lie on the vertices of the
Dynkin diagram of the Lie algebra $D_6$, as it will turn out. This model is exactly
solvable, both in the original, as well as the Galois conjugated case.

\begin{figure}[!hb]
\begin{center}
 \includegraphics[width=.75\columnwidth]{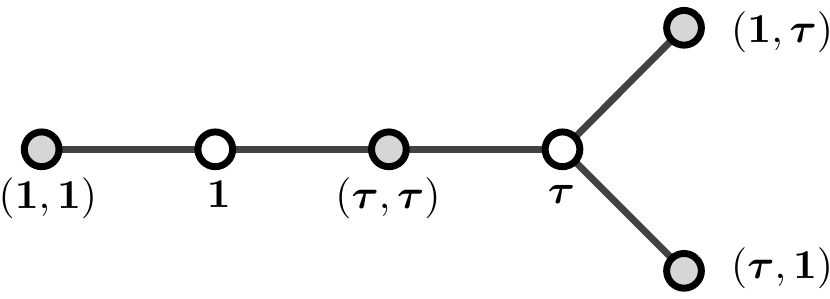}
\end{center}
\caption{The allowed `height' configurations of the ladder models. This diagram
is the Dynkin diagram of the Lie algebra $D_6=so(12)$.}
\label{Fig:D6diagram}
\end{figure}

\begin{widetext}
The basis transformation: At each rung, one applies an $F$-transformation,
changing the basis building blocks of the ladder:
\begin{figure}[h]
 \begin{center}
 \includegraphics[width=.9\columnwidth]{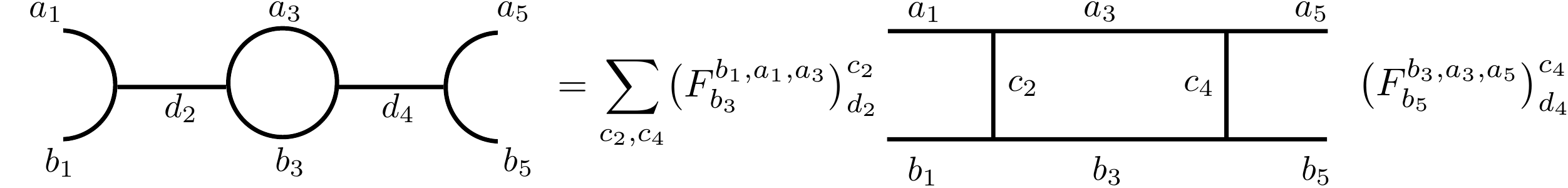}
 \end{center}
\caption{Basis transformation of the ladder model.}
\label{Fig:laddertransform}
\end{figure}
\end{widetext}
Thus, we can express the Hilbert space in terms of the variables
$\{a_{2i-1},b_{2i-1},c_{2i} \}$, or, equivalently, in terms of the
variables
$\{a_{2i-1},b_{2i-1},d_{2i} \}$, as displayed in figure \ref{Fig:laddertransform}.
The mapping to the restricted solid-on-solid model is performed in terms of the
latter variables.
In particular, one can determine which `values' the links can take on the even and
odd sublattices. The variables on the even sublattice $d_{2i}$
can take the values $\{ \id, \btau \}$, while
the combined variables $(a_{2i-1},b_{2i-1})$ can take the values
$\{ (\id,\id), (\id,\btau), (\btau,\id), (\btau,\btau) \}$, but there are constraints on which values of
$(a_{2i-1},b_{2i-1})$ and $d_{2i}$ can occur next to each other. From the fusion rules,
it easily follows that the combination of variables which are consistent with one another,
have to be adjacent on the `height graph', as shown in Fig.~\ref{Fig:D6diagram}.

From now on, we will call the allowed variables (or pairs of variables on the
odd sublattice), heights. The Hilbert space of the ladder model consists of all
height configurations on a chain of length $L$, in such a way that two heights
on neighboring sites also have to be neighbors (i.e., be connected by a solid
line) in the height diagram of Fig.~\ref{Fig:D6diagram}. In this way, the model can
be mapped on the $D_6$ version of the restricted solid on solid model,
and precisely when $J_r = J_p = \pm 1$, the model is critical, and can be
solved exactly. We will not go into the details here (which can be found in
Ref.~\onlinecite{Gils09b}), but we will quickly state the result for this critical behavior
in the original, Fibonacci anyon setting, followed by the results on the
Galois conjugated Yang-Lee ladder. The details of the phase diagram,
which closely mimics the phase diagram of the
unitary (`not Galois-conjugated') ladder, and the
numerical results will be presented in the next subsection.

As was the case for the chain models, also in the ladder case do the
parts of the Hamiltonian form a representation of the Temperley-Lieb
algebra, but in this case, the algebra is 
that\cite{PasquierNPB1987} based on the Dynkin diagram $D_6$.
The possible associated values of the
$d$-isotopy parameter are given by $d=2\cos(\frac{m_j \pi}{h})$, where
$h=10$, and the $m_j$ are the exponents of the Lie algebra $D_6$,
namely $m_j = 1,3,5,7,9,5$. In the original ladder model $m_j=1$, for
both cases $J_r=J_p = \pm 1$. The associated value of the $d$-isotopy
reads $d=2\cos(\pi/10)=\sqrt{2+\phi}$. In the antiferromagnetic case, i.e.
$J_r = J_p = 1$, or in the parametrization $J_p=\cos\theta$, $J_r=\sin\theta$,
the point $\theta = \pi/4$, the critical behavior is described by the
minimal model $\widetilde{\mathcal{M}}(9,10)$, which has central
charge $c=14/15$. With the tilde, we denote that we are considering
the $(A_8,D_6)$
 modular invariant\cite{CardyModularInvariants,ciz87a}, instead 
of the 
`usual' (diagonal)
 $(A_8,A_9)$ invariant
(we note that the $(A_8,D_6)$ invariant minimal model $\widetilde{\mathcal{M}}(9,10)$
is equivalent to the coset model $(G_2)_1 \times (G_2)_1/(G_2)_2$),
where
$(G_2)_k$ denotes the Wess-Zumino-Witten CFT at level k based on
the Lie-algebra $G_2$.

We will now switch our attention to the Galois
conjugated version of the ladder model.
The phase diagram of the Yang-Lee ladder is depicted in
figure \ref{Fig:PhaseDiagram-DoubledChain}, and the location of the
gapped phases, as well as the critical behavior is exactly the same as in
the Fibonacci ladder model. Therefore, we will be somewhat brief in the
description of these phases (see the introduction of this section for a
brief description of the gapped phases), and focus on the differences,
which occur at the exactly solvable points, namely $\theta = \pi/4,5\pi/4$,
where the model is critical, but with a different critical behavior in
comparison to the Fibonacci ladder.

\paragraph*{The antiferromagnetic $(\theta=\pi/4)$ critical point.--}

As was the case in the Yang-Lee chains, the critical behavior
of the Galois conjugated model is obtained by picking a different
$d$-isotopy parameter, corresponding to the Galois conjugate
under consideration. For Fibonacci anyons, there is only one
Galois conjugate, but nevertheless, there seem to be different
possible values for the $d$-isotopy parameter, namely
$d=2\cos(\frac{m_j \pi}{h})$, with
$h=10$, and $m_j = 1,3,5,7,9,5$.  In fact, the value
we need is $m_j=3$, which corresponds to
the value obtained from the Fibonacci case, under Galois conjugation,
namely $d=\sqrt{2-1/\phi} = 2 \cos(3\pi/10)$. 

It is a simple matter to check that this value corresponds to a model
with central charge $c = 8/35$, by using the same arguments as we gave
in section \ref{Sec:su2k}. This central charge corresponds to the central
charge of the non-unitary minimal model $\mathcal{M}(7,10)$. 
Just as in the unitary case\cite{Gils09b},
 the critical CFT describing the Yang-Lee model at
$\theta = \pi/4$, is not the usual (`diagonal') modular invariant, but rather the
$(A,D)$ invariant, or, to be more precise, $(A_6,D_6)$. See \cite{ciz87a}
for information on the different modular invariants of the minimal models.
We will denote this model by $\widetilde{\mathcal{M}}(7,10)$.
The field content of these models is given in tables \ref{Tab:KacM710310}
and \ref{Tab:KacM710310Dinv} below. Details of the spectrum of this model will
be discussed in the section \ref{Sec:YL-ladder-num} below.

\begin{table}[t]
  \begin{tabular}{c|c|c|c}
     \hline
     \hline
     \multicolumn{4}{c}{\bf Fibonacci ladder} \\
     \hline
     coupling & CFT & central charge & $d$-isotopy \\
     \hline
     AFM $(\theta=\pi/4)$ & $\mathcal{M}(9,10)$ & $c = 14/15$ & $2\cos{\left( \frac{\pi}{10} \right)}$ \\
     FM   $(\theta=5\pi/4)$ & $\mathcal{Z}_8$ parafermion & $c=7/5$ & $2\cos{\left( \frac{\pi}{10} \right)}$ \\
     \hline
     \multicolumn{4}{c}{\bf Yang-Lee ladder}\\
     \hline
     coupling & CFT & central charge & $d$-isotopy \\
     \hline
     AFM $(\theta=\pi/4)$ & $\widetilde{\mathcal{M}}(7,10)$ & $c = 8/35$ & $2\cos{\left( \frac{3\pi}{10} \right)}$ \\
     FM   $(\theta=5\pi/4)$ & $\widetilde{\mathcal{M}}(3,10)$ & $c = -44/5$ & $2\cos{\left( \frac{7\pi}{10} \right)}$ \\
     \hline
     \hline
  \end{tabular}
  \caption{Gapless theories for the ladder models.
  }
  \label{Tab:Critical-ladder}
\end{table}

\paragraph*{The ferromagnetic $(\theta=5\pi/4)$ critical point.--}

In the non-unitary case, we can obtain the critical theory for the opposite
sign of the interaction, by simply swapping the sign of the $d$-isotopy
parameter, which hence corresponds to the value
$d=2\cos(7\pi/10)=-\sqrt{2-1/\phi}$, from which it is a simple matter to
extract the central charge, which is given by $c=-44/5$, corresponding to
the minimal model $\mathcal{M}(3,10)$. This minimal model also comes in
two incarnations, one corresponding to the modular invariant
$(A_2,A_9)$, the other to $(A_2,D_6)$.
(The field content of these models is also given in tables \ref{Tab:KacM710310}
and \ref{Tab:KacM710310Dinv} below.) The latter is
realized in the Yang-Lee ladder at $\theta = 5\pi/4$, and we will describe
the spectrum in more detail below. Before we do that, however, we
will first summarize the critical theories of both the Fibonacci and Yang-Lee
anyon ladders in table \ref{Tab:Critical-ladder}.

\begin{table}
  \begin{tabular}{r|ccc}
  \hline \hline
  \multicolumn{4}{c}{$\mathcal{M}(7,10)$} \\
  \hline 
  $h(r,s)$ & $s=1$ & $3$ & $5$ \\
  \hline
  $r=1$ & $0$ & $13/7$ & $46/7$ \\
  $2$ & $1/40$ & $247/280$ & $1287/280$ \\
  $3$ & $2/5$ & $9/35$ & $104/35$ \\
  $4$ & $9/8$ & $-1/56$ & $95/56$ \\
  $5$ & $11/5$ & $2/35$ & $27/35$ \\
  $6$ & $29/8$ & $27/56$ & $11/56$ \\
  $7$ & $27/5$ & $44/35$ & $-1/35$ \\
  $8$ & $301/40$ & $667/280$ & $27/280$ \\
  $9$ & $10$ & $27/7$ & $4/7$ \\
  \hline \hline
  \end{tabular}
  \hskip 8mm
  \begin{tabular}{r|c}
  \hline \hline
  \multicolumn{2}{c}{$\mathcal{M}(3,10)$} \\
  \hline 
  $h(r,s)$ & $s=1$\\
  \hline
  $r=1$ & $0$ \\
  $2$ & $-11/40$ \\
  $3$ & $-2/5$ \\
  $4$ & $-3/8$ \\
  $5$ & $-1/5$ \\
  $6$ & $1/8$ \\
  $7$ & $3/5$ \\
  $8$ & $49/40$ \\
  $9$ & $2$ \\
  \hline \hline
  \end{tabular}
  \caption{Kac table for the non-unitary minimal models $\mathcal{M}(7,10)$ and
  $\mathcal{M}(3,10)$. Only odd values
  of $r$ are displayed, so that each field appears only once.}
  \label{Tab:KacM710310}
\end{table}

\begin{table}[h]
  \begin{tabular}{r|ccc}
  \hline \hline
  \multicolumn{4}{c}{$\widetilde{\mathcal{M}}(7,10)$} \\
  \hline 
  $h(r,s)$ & $s=1$ & $3$ & $5$ \\
  \hline
  $r=1$ & $0$ & $13/7$ & $4/7$ \\
  $3$ & $2/5$ & $9/35$ & $-1/35$ \\
  $5$ & ${\bf 11/5}$ & ${\bf 2/35}$ & ${\bf 27/35}$ \\
  \hline \hline
  \end{tabular}
  \hskip 8mm
  \begin{tabular}{r|c}
  \hline \hline
  \multicolumn{2}{c}{$\widetilde{\mathcal{M}}(3,10)$} \\
  \hline 
  $h(r,s)$ & $s=1$\\
  \hline
  $r=1$ & $0$ \\
  $3$ & $-2/5$ \\
  $5$ & ${\bf -1/5}$ \\
  \hline \hline
  \end{tabular}
  \caption{Kac table for the `$(A,D)$-modular invariant' non-unitary minimal models
  $\widetilde{\mathcal{M}}(7,10)$ and
  $\widetilde{\mathcal{M}}(3,10)$. The fields with the conformal dimensions
  in bold (i.e. those with label $s=5$) appear twice.}
  \label{Tab:KacM710310Dinv}
\end{table}


\subsection{Numerical results}
\label{Sec:YL-ladder-num}

\begin{figure}[t]
\begin{center}
  \includegraphics[width=\columnwidth]{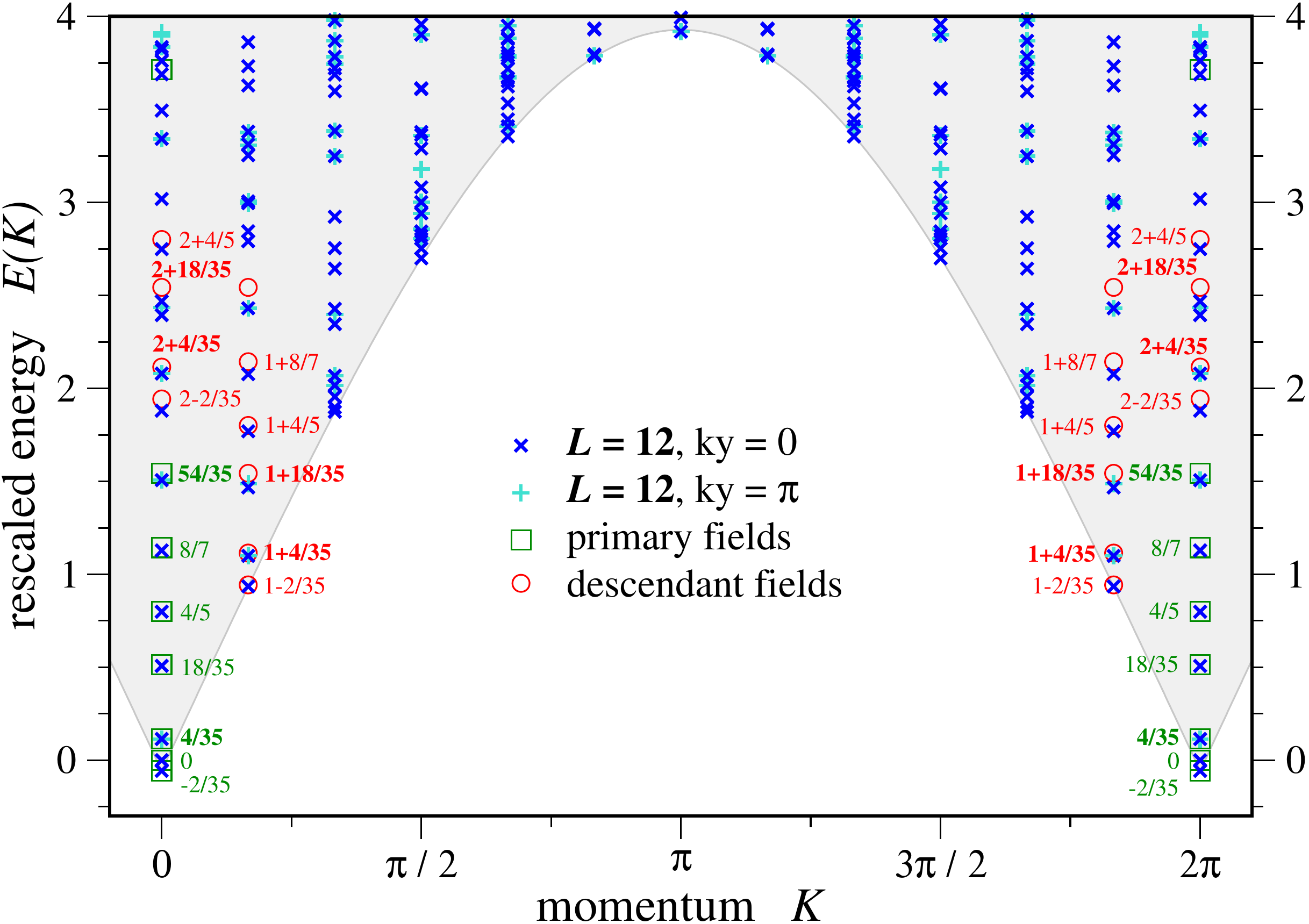}
\end{center}
\caption{
   Conformal energy spectrum of the critical points in the doubled `antiferromagnetic' Yang-Lee chain.
   The spectrum matches the non-unitary minimal model $\widetilde{M}(7,10)$ 
   with central charge $c=8/35$.
   Primary fields of the conformal field theory are indicated by green squares, descendant fields by
   red circles.
 }
\label{Fig:AFM-DoubledChain}
\end{figure}

We finally present the numerical spectra of the conjugated ladder model at the
two critical points discussed in the previous section.
The spectrum for the critical point at $\theta = \pi/4$, is given in
Fig.~\ref{Fig:AFM-DoubledChain}, where we indicated the locations of the
primary fields of $\widetilde{\mathcal{M}}(7,10)$ (given in table \ref{Tab:KacM710310Dinv})
by green squares, as well as some low-lying descendants with red circles.
As usual, there are only two free parameters to match the numerically
obtained spectrum with the result obtained from conformal field
theory, so the fact that the six lowest primaries, as well as several
descendants match to high precision (limited by finite size effects) is a
very non-trivial check on our results. 

\begin{figure}[t]
\begin{center}
  \includegraphics[width=\columnwidth]{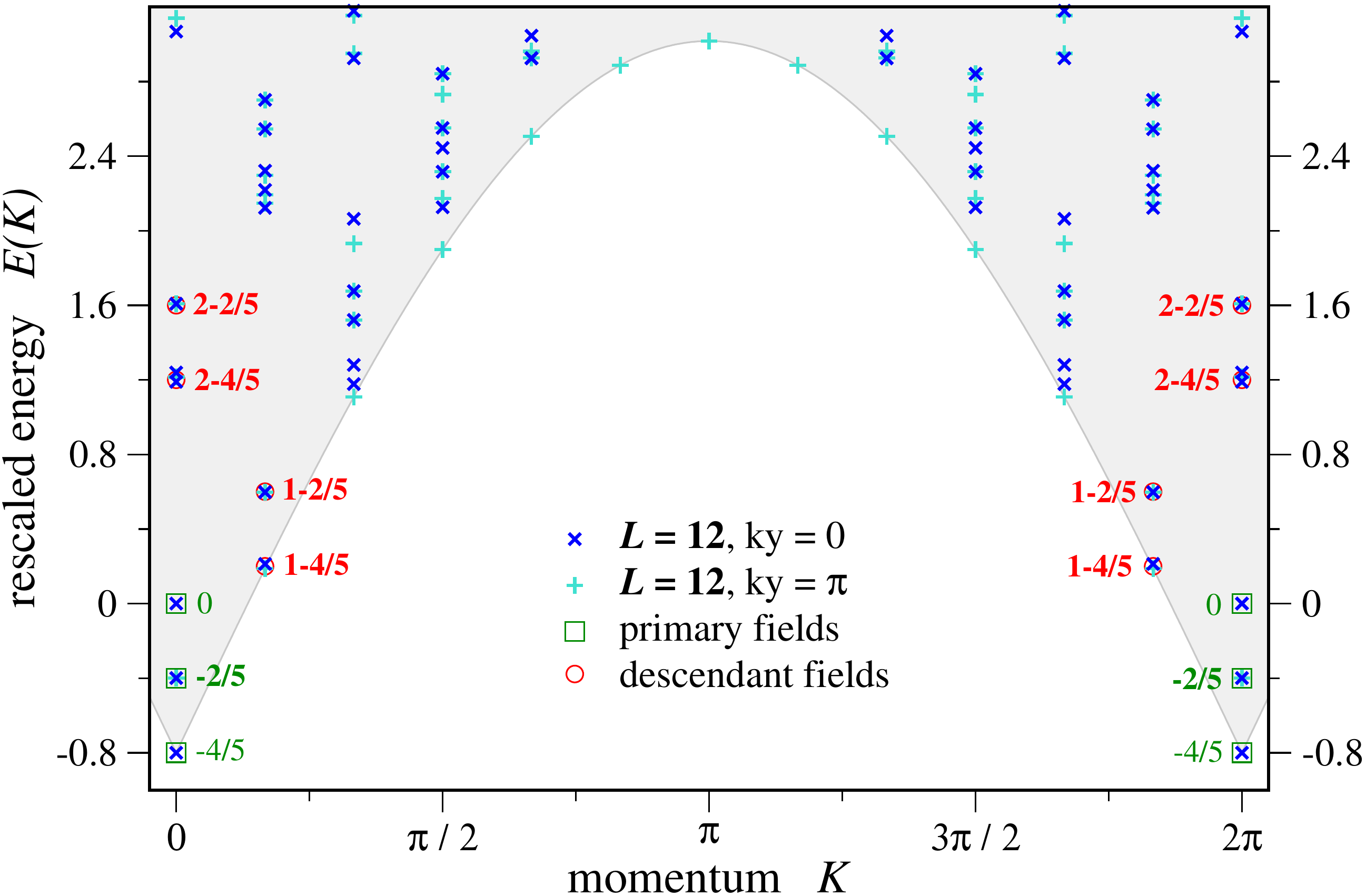}
\end{center}
\caption{
   Conformal energy spectrum of the critical points in the doubled `ferromagnetic' Yang-Lee chain.
   The spectrum matches the non-unitary minimal model $\widetilde{M}(3,10)$ 
   with central charge $c=-44/5$.
   Primary fields of the conformal field theory are indicated by green squares, descendant fields by
   red circles.
 }
\label{Fig:FM-DoubledChain}
\end{figure}

In Figure \ref{Fig:FM-DoubledChain}, we give the numerical spectrum of the
ferromagnetic Yang-Lee ladder, at $\theta = 5\pi/4$, which is characteristic
of the critical phase extending over $\theta \in (\pi,3\pi/2)$. In this case the
critical behavior is described by the $\widetilde{\mathcal{M}}(3,10)$ non-unitary
conformal field theory, and, as for the antiferromagnetic case, we were able
to identify the primary fields, as well as several low-lying descendant fields,
as indicated in the figure. The fields of the CFT are given in table
\ref{Tab:KacM710310Dinv}.

\section{Discussion \& Summary}
\label{Sec:Summary}

In this paper, we studied the collective states of Yang-Lee anyons, a family of non-unitary, non-Abelian anyons
which are close cousins of the unitary Fibonacci anyons. Non-unitary anyons of this form have attracted
interest in the context of studies of certain quantum Hall wavefunctions, including the Gaffnian state \cite{Gaffnian}.
Both Yang-Lee and Fibonacci anyons arise from the same anyonic theory, su(2)$_3$, and in particular 
they share the same fusion rules. The key distinction between the two anyon types is that Yang-Lee anyons are
non-unitary and relate to their unitary counterparts, the Fibonacci anyons, via `Galois conjugation'.
We have generalized this concept to arbitrary su(2)$_k$ anyonic theories.

To characterize the collective states formed by a set of anyons in the presence of pairwise interactions, 
we have considered one-dimensional models of interacting Yang-Lee anyons similar to the golden chain 
model of the unitary case \cite{Feiguin07}.
Analogous to the case of interacting Fibonacci anyons, the collective states of such chains of Yang-Lee anyons 
are found to be critical and  the gapless theories are described by certain non-unitary 
conformal field theories (which depend on the sign of the coupling, see Table \ref{Tab:Chainresults}).

The non-unitary chain models are found to exhibit rather peculiar features related to the presence of a non-local, 
topological symmetry first observed in the unitary models. For the multicritical gapless theories of these chain 
models this topological symmetry, whose related symmetry operator commutes with the Hamiltonian, 
allows to classify all operators by a topology symmetry sector. It turns out that all relevant operators corresponding
to (uniform) perturbations of the gapless system are in a different symmetry sector from the one of the ground state.
In the unitary case, this mechanism which effectively protects the gapless ground state from local perturbations has 
led to an interpretation \cite{Gils09} of these gapless modes of a chain of interacting anyons as {\sl edge states} of 
the parent topological liquid of which the anyons are excitations of. 
This has further led to the conclusion that interactions between anyons (in two-dimensional arrangements) result 
in a splitting of the macroscopic degeneracy of a set of non-Abelian anyons and the nucleation of a new topological 
liquid inside and spatially separated from the parent liquid of which the anyons are excitations \cite{Gils09,Ludwig10}.

For the non-unitary models, studied in this manuscript, a crucial distinction comes in the symmetry sector assigned
to the ground state. While in the unitary case, the ground state was found to be in the flux-free (topological trivial) sector, 
we find that the ground state in the non-unitary case exhibits a non-trivial topological flux (i.e. is in the non-trivial topological 
symmetry sector). 
It remains an open question whether such a spontaneous creation of topological flux can occur in the ground state of a system
of interacting anyonic quasiparticles residing in a bulk-gapped topological liquid, which has no topological flux associated 
with it. This may be an indication that the non-unitary anyons are not massive quasiparticles of a gapped quantum liquid, but in fact 
excitations of a gapless quantum liquid.

In summary, we have investigated one-dimensional models of 
non-unitary anyons based on su(2)$_k$ anyonic theories. 
The collective ground states are found to display critical behavior, similar to their 
unitary counterparts to which they are related via a `Galois conjugation'. 
We described in detail the non-unitary conformal field theories capturing their critical behavior
and commented on possible physical implications of our results in connection with
proposed quantum Hall states related to such non-unitary CFTs. 
In addition, we also studied the phase diagram of an interacting ladder model, which is a 
quasi one-dimensional version of the Levin-Wen model for non-unitary Yang-Lee anyons.


\begin{acknowledgments}
We acknowledge discussions with P.~Bonderson, M.~Freedman, J.~Slingerland, and Z.~Wang.
A.W.W.L. was supported, in part, by NSF DMR-0706140. 
We thank the Aspen Center for Physics and the Kavli Institute for Theoretical Physics supported
by NSF PHY-0551164.
Our numerical work used some of the ALPS libraries, \cite{alps20,alps13} see also http://alps.comp-phys.org.
\end{acknowledgments}

\appendix

\section{Detailed description of the conformal energy spectra.}

In this section, we will describe in some detail the structure of the descendant fields
present in the numerically obtained spectra. In the case of the Yang-Lee ladders, this
structure is somewhat different from the `usual' structure. The object containing the
information about the number of states at a specific energy and momentum (in the
thermodynamic limit) is the partition function, which we will denote by $Z_{\rm tot}$.

Modular invariance of the partition function on the torus constraints the possible
partition functions of rational conformal field theories at a particular central
charge \cite{ciz87a}. In this paper, we will only be concerned with the partition functions
of minimal models, which are described in terms of the chiral characters
associated to the primary fields. An explicit expression for these 
chiral characters associated to the primary fields
$\phi_{(r,s)}$ of the minimal models $\mathcal{M}(p,p')$ is given in,
for instance, chapter 8 of reference \onlinecite{book:fms99}.

\begin{equation}
{\rm ch}_{(r,s)}^{(p,p')} (q)= \frac{q^{h_{(r,s)}}}{(q)_\infty}
\sum_{k \in \mathbb{Z}} \left( q^{k (k p p' + r p - s p')}  - q^{(k p + s) (k p' + r)} \right) \ ,
\end{equation}
where $(q)_\infty = \prod_{k>0} (1-q^k)$, and
$h_{(r,s)} = \frac{(pr-p's)^2-(p-p')^2}{4pp'} $
the conformal dimension of the field $\phi_{(r,s)}$.

Dropping the labels $(p,p')$, all  the possible partition functions can be written
as
\begin{equation}
Z_{\rm tot} = \sum_{(r,s);(r',s')} M_{(r,s);(r',s')}\ch_{(r,s)} \ch^*_{(r',s')} \ .
\end{equation}
For the minimal models, there is always the so-called `diagonal modular
invariant', for which $M$ is diagonal, $M_{(r,s);(r',s')} = \delta_{r,r'}\delta_{s,s'}$,
and the sum in the partition function runs over all primary fields. However, in
general, there exist other modular invariant partition functions. The critical points
of the ladder models realize some of these, as we pointed out in the main text.

In terms of the Virasoro generators $L_0$ and $\bar{L}_0$, the Hamiltonian and
momentum operators can be written (after an appropriate rescaling and shift) as
$H = L_{0} + \bar{L}_{0}$ and $P = (L_{0}-\bar{L}_{0})$. The partition function can
be written as a trace over the Hilbertspace,
$Z_{\rm tot} = {\rm tr} \; q^{L_{0}}\bar{q}^{\bar{L}_{0}}$,
which allows us to extract both the energies and momenta for each state.

In particular, the energy of a state is given by the sum of the powers of $q$ and
$\bar{q}$. We note that this energy is measured with respect to the energy
corresponding to the energy of the identity operator ${\bf 1}$, which we have set
to zero. In addition, the momenta of a state is given by $L_{0}-\bar{L}_{0}$, in
units of $2\pi/L$, and measured with respect to the momenta of the
corresponding primary field, which is not determined by conformal field theory,
but rather by the specific Hamiltonian realization.

\subsection{The ferromagnetic Yang-Lee ladder}

The critical theory describing the ferromagnetic Yang-Lee ladder is the
$(D_6,A_2)$ modular invariant $\widetilde{\mathcal{M}}(3,10)$.
The total partition function of this theory can be written as follows\cite{ciz87a}
(we will drop the labels $(p,p')=(3,10)$ from the characters for convenience)
\begin{align}
Z &=
|\ch_{(1,1)}|^2 + |\ch_{(3,1)}|^2 + 2 |\ch_{(5,1)}|^2 + |\ch_{(7,1)}|^2 \\
& + |\ch_{(9,1)}|^2 + (\ch_{(1,1)} \ch_{(1,2)}^*+ \ch_{(3,1)} \ch_{(3,2)}^* + {\rm c.c.}) \ .
\nonumber
\end{align}

By expanding the partition function, in terms of $q$ and $\bar{q}$, we can
completely explain the low-lying part of the spectrum displayed in figure
\ref{Fig:FM-DoubledChain}. Let us start with the vacuum sector of the theory,
which corresponds to the part of the partition function with integer powers
of $q$ and $\bar{q}$:
\begin{align}
Z_{0} &= |\ch_{(1,1)}|^2 + |\ch_{(9,1)}|^2 +  (\ch_{(1,1)} \ch_{(1,2)}^*+ {\rm c.c.}) =
\nonumber  \\
&1 + 2q^2 + 2\bar{q}^2 + \cdots \ . 
\end{align}
The total power of $q$ and $\bar{q}$ gives the energy of the states, while
the difference between the powers gives the momentum. Thus, the character
of the vacuum sector implies the presence of two states at energy $E=2$ and $K_x = 2$
as well as two states at energy $E=2$ and momentum $K_x =-2$. Note that we count the
states irrespective of their $K_y$ sector.

We will continue with the descendants corresponding to the primary field
with energy $E=-2/5$ and momentum $K_x = 0$. At this energy and momentum,
there are in fact two states. This sector of the partition function reads
\begin{align}
&Z_{-2/5} = 2 |\ch_{(5,1)}|^2 = \nonumber \\ 
&2q^{-1/5} \bar{q}^{-1/5} \left(
1 + q + \bar{q} + 2q^2 + 2 \bar{q}^2 + q \bar{q}+
\cdots \right) \ , 
\end{align}
which implies two states at $(E,K_x) = (1-2/5,1)$ and two at
$(E,K_x) = (1-2/5,-1)$, as well as two states at $(2-2/5,0)$ and four
at both $(2-2/5,2)$ and $(2-2/5,-2)$, all of which is reproduced (modulo
finite size effects) in the spectrum in figure \ref{Fig:FM-DoubledChain}.

Finally, we consider the descendants corresponding to the primary field at $E=-4/5$,
which also has momentum $K_x = 0$.
This sector of the partition function reads
\begin{align}
&Z_{-4/5} = |\ch_{(3,1)}|^2 + |\ch_{(7,1)}|^2 + (\ch_{(3,1)} \ch_{(3,2)}^*+ {\rm c.c.})  = \nonumber \\ 
&q^{-2/5} \bar{q}^{-2/5} \left(
1+ 2 q + 2 \bar{q} + 3q^2+3\bar{q}^3+ 4 q \bar{q} +  
\cdots \right) \ .
\end{align}
This implies the presence of two states at both $(E,K_x) = (1-4/5,1)$ and
$(1-4/5,1)$, as well as three states at both $(2-4/5,2)$ and
$(2-4/5,-2)$. In addition, there should be four states at $(2-4/5,0)$. This seems
to be at odds with the figure \ref{Fig:FM-DoubledChain}, but it turns out that the
state at $K_x = 0$ and $K_y = \pi$, and energy of approximately $E\approx 2-4/5$ (or,
more precisely, $E \approx 1.2144495$) is in fact doubly degenerate, so indeed, there
are four states, as expected from the partition function.

\subsection{The antiferromagnetic Yang-Lee ladder}

A similar analysis can be performed for the antiferromagnetic critical point
of the Yang-Lee ladder. We will be brief here, because the analysis is
identical to the one for the ferromagnetic critical point.

There are nine sectors in this particular case, corresponding to the fields
displayed in table \ref{Tab:KacM710310Dinv}, so the partition function
can be written as
\begin{equation}
\begin{split}
Z_{\rm tot} &= Z_{0} + Z_{4/5} + Z_{22/5} + Z_{26/7} + Z_{18/35} + Z_{4/35}\\
&+ Z_{8/7} + Z_{-2/35} + Z_{54/35}
\end{split}
\end{equation}
Although we will not consider all of these sectors (some give rise to states which
are too high in energy to be unambiguously identified, due to finite size effects),
we will nevertheless give the form of the partition function restricted to each of
these sectors. We will drop the labels $(p,p')=(7,10)$ from the characters
\begin{align*}
Z_{0} &= |\ch_{(1,1)}|^2 + |\ch_{(9,1)}|^2 + (\ch_{(1,1)} \ch_{(1,6)}^* + {\rm c.c.}) \\
Z_{4/5} &= |\ch_{(3,1)}|^2 + |\ch_{(7,1)}|^2 + (\ch_{(3,1)} \ch_{(3,6)}^* + {\rm c.c.}) \\
Z_{22/5} &= 2 |\ch_{(5,1)}|^2 \\
Z_{26/7} &= |\ch_{(1,3)}|^2 + |\ch_{(9,3)}|^2 + (\ch_{(1,3)} \ch_{(1,4)}^* + {\rm c.c.}) \\
Z_{18/35} &= |\ch_{(3,3)}|^2 + |\ch_{(7,3)}|^2 + (\ch_{(3,3)} \ch_{(3,4)}^* + {\rm c.c.}) \\
Z_{4/35} &= 2 |\ch_{(5,3)}|^2 \\
Z_{8/7} &= |\ch_{(1,5)}|^2 + |\ch_{(9,5)}|^2 + (\ch_{(1,5)} \ch_{(1,2)}^* + {\rm c.c.}) \\
Z_{-2/35} &= |\ch_{(3,5)}|^2 + |\ch_{(7,5)}|^2 + (\ch_{(3,5)} \ch_{(3,2)}^* + {\rm c.c.}) \\
Z_{54/35} &= 2 |\ch_{(5,5)}|^2
\end{align*}

The parts of the partition function read explicitly
\begin{align*}
Z_{0} &= 1+ q^2 + \bar{q}^2 +\ldots \\
Z_{4/5} &= q^{2/5}\bar{q}^{2/5}\left(1+q+\bar{q}+2q^2+q\bar{q}+2\bar{q}^2 +\ldots \right) \\
Z_{22/5} &= 2q^{11/5}\bar{q}^{11/5}\left(1+q+\bar{q}+2q^2+q\bar{q}+2\bar{q}^2 +\ldots \right) \\
Z_{26/7} &= q^{13/7}\bar{q}^{13/7}\left(1+q+\bar{q}+3q^2+q\bar{q}+3\bar{q}^2 +\ldots \right) \\
Z_{18/35} &= q^{9/35}\bar{q}^{9/35}\left(1+2q+2\bar{q}+3q^2+4q\bar{q}+3\bar{q}^2 +\ldots \right) \\
Z_{4/35} &= 2q^{2/35}\bar{q}^{2/35}\left(1+q+\bar{q}+2q^2+q\bar{q}+2\bar{q}^2 +\ldots \right) \\
Z_{8/7} &= q^{4/7}\bar{q}^{4/7}\left(1+q+\bar{q}+q^2+q\bar{q}+\bar{q}^2 +\ldots \right) \\
Z_{-2/35} &= q^{-1/35}\bar{q}^{-1/35}\left(1+q+\bar{q}+2q^2+q\bar{q}+2\bar{q}^2 +\ldots \right) \\
Z_{54/35} &= 2q^{27/35}\bar{q}^{27/35}\left(1+q+\bar{q}+2q^2+q\bar{q}+2\bar{q}^2 +\ldots \right) 
\end{align*}

With this information, it is rather straight forward to check that the primaries and descendant
fields we indicated in figure \ref{Fig:AFM-DoubledChain} indeed come with the right multiplicities.

\subsection{The ferromagnetic Yang-Lee chain}

After having dealt with the (anti)ferromagnetic Yang-Lee ladders in quite some detail,
we will content ourselves here by giving the (diagonal) partition functions describing
the spectra, and note that for all states where finite size effects allow us to make an
identification, we obtain full agreement with the CFT prediction.

The partition function in case of the ferromagnetic Yang-Lee chain, the critical model
is given by the diagonal invariant of the model $\mathcal{M}(2,5)$, whose partition
function reads (dropping the label $(2,5)$ on the chiral character)
\begin{equation}
Z = |\ch_{(1,1)}|^2 + |\ch_{(3,1)}|^2 = Z_{0} + Z_{-2/5} \ ,
\end{equation}
with
\begin{align}
Z_{0} &= 1+ q^2 + \bar{q}^2 + q^3 + \bar{q}^3 + q^4 + q^2 \bar{q}^2 + \bar{q}^4 + \nonumber\\
& q^5 + q^3 \bar{q}^2 + q^2 \bar{q}^3 + \bar{q}^5 +\nonumber\\
& 2q^6 + q^4 \bar{q}^2 + q^3 \bar{q}^3 + q^2 \bar{q}^4 + 2\bar{q}^6 + \cdots \\ 
Z_{-2/5} &= q^{-1/5} \bar{q}^{-1/5}\bigl(1 + q + \bar{q} + q^2 + q\bar{q} + \bar{q}^2 +\\
& q^3 + q^2 \bar{q} + q \bar{q}^2 + \bar{q}^3 + \nonumber \\
& 2 q^4 + q^3 \bar{q} + q^2 \bar{q}^2 + q \bar{q}^2 + 2\bar{q}^4 + \nonumber \\
& 2q^5 + 2q^4 \bar{q} + q^2 \bar{q}^3 + q^3 \bar{q}^2 + 2q \bar{q}^4 + 2\bar{q}^5 + \nonumber \\
& 3q^6 + 2q^5 \bar{q} + 2q^4 \bar{q}^2 + q^3 \bar{q}^3
+ 2q^2 \bar{q}^4 + 2q \bar{q}^5 + 3\bar{q}^6 \nonumber \\
&+ \cdots \bigr) \nonumber 
\end{align}
All these states, and in fact quite a few more which we did not give here, are
reproduced in the spectrum of the ferromagnetic Yang-Lee chain, as shown
in figure \ref{Fig:FM-Chain}.

\subsection{The antiferromagnetic Yang-Lee chain}

Finally, we deal with the antiferromagnetic Yang-Lee chain, the critical behavior
of which is described by the (diagonal invariant of the) minimal model $\mathcal{M}(3,5)$,
which has four primary fields, as described in the main text. The relevant partition function
is given by (again, dropping the labels $(3,5)$ on the chiral characters)
\begin{align}
Z &= |\ch_{(1,1)}|^2 + |\ch_{(3,1)}|^2 + |\ch_{(1,2)}|^2 + |\ch_{(3,2)}|^2 \nonumber \\
&= Z_{0} + Z_{2/5} + Z_{3/2} + Z_{-1/10}   \ ,
\end{align}
with
\begin{align}
Z_{0} &= 1+ q^2 + \bar{q}^2 + \cdots \\
\nonumber 
Z_{2/5} &=  q^{1/5} \bar{q}^{1/5} \bigl(1 + q + \bar{q} + 2 q^2 + q \bar{q} + 2\bar{q}^2 + \cdots \bigr) \\
\nonumber 
Z_{3/2} &=  q^{-3/4} \bar{q}^{-3/4} \bigl(1 + q + \bar{q} + \cdots \bigr) \\
\nonumber 
Z_{-1/10} &=  q^{1/20} \bar{q}^{1/20} \bigl(1 + q + \bar{q} + q^2 + q \bar{q} + \bar{q}^2 + \\
\nonumber
& + 2 q^3 + q^2 \bar{q} + q \bar{q}^2 + 2 \bar{q}^3 \cdots \bigr) 
\end{align}
We should note that, as can be seen in figure \ref{Fig:AFM-Chain}, these fields
are reproduced, and that (for $L = 32$ sites), the fields with conformal dimension
$h = -1/20$ and $h = 3/4$ occur at momentum $K_x = 0$, while the identity fields and
the field with conformal dimension $h = 1/5$ occur at momentum $K_x = \pi$.



\begin{thebibliography}{100}
\bibitem{FootnoteGauge}
We note that in a different gauge, the matrix $F$-matrix \eqref{Eq:Fmatgen}
takes the form
$F^{\btau,\btau,\btau}_{\btau}=\begin{pmatrix}
\frac{1}{q}+ q - 1 &\frac{1}{q}+ q  + 1 \\ 2 (\frac{1}{q}+ q) - 3 & -(\frac{1}{q}+ q) + 1
\end{pmatrix}
$. This $F$-matrix leads to different forms of the Fibonacci and Yang-Lee Hamiltonians as
given in Eq.~\eqref{Eq:HMatrix}, but they nevertheless have identical energy spectra.
\bibitem{FootnoteVariant}
The minimal model variants $\widetilde{\mathcal{M}}(p,p')$ are discussed in more detail
in section \ref{Sec:PhaseDiag}.
\bibitem{FootnoteIntegerRepresentations}
More generally, one finds that the integer representations form
a closed subset under the fusion rules of Eq.~\eqref{Eq:su2kfusion}.
\bibitem{FootnoteNonUnitary}
The identification of the critical theories between regions III and IV in the Forrester-Baxter RSOS model \cite{fb84} with these non-unitary minimal models was found in Refs.~\onlinecite{Riggs} and \onlinecite{Nakanishi}.
\end{thebibliography}
\end{document}